%% file: anar_paper.tex
\documentclass[sigchi]{acmart}

\usepackage{balance}
\usepackage{booktabs} 
\copyrightyear{2019}
\acmYear{2019}
\setcopyright{rightsretained}
\acmConference[CHI 2019]{CHI Conference on Human Factors in Computing Systems Proceedings}{May 4--9, 2019}{Glasgow, Scotland, UK}
\acmBooktitle{CHI Conference on Human Factors in Computing Systems Proceedings (CHI 2019), May 4--9, 2019, Glasgow, Scotland, UK}
\acmDOI{10.1145/3290605.3300569}
\acmISBN{978-1-4503-5970-2/19/05}

\begin{document}
\settopmatter{printacmref=true}
\fancyhead{}
\title{Human-Computer Insurrection}
\subtitle{Notes on an Anarchist HCI}

\author{Os Keyes}
\orcid{0000-0001-5196-609X}
\affiliation{%
  \institution{University of Washington}
  \streetaddress{428 Sieg Hall, Box 352315}
  \city{Seattle}
  \state{WA}
  \postcode{98195-2315}
  \country{USA}}
\email{okeyes@uw.edu}
\authornote{All authors contributed equally to this work and consider it a collective creation.  Intellectual property is theft.}

\author{Josephine Hoy}
\affiliation{%
  \institution{University of Washington}
  \streetaddress{428 Sieg Hall, Box 352315}
  \city{Seattle}
  \state{WA}
  \postcode{98195-2315}
  \country{USA}}
\email{joeyhoy@uw.edu}
\authornotemark[1]

\author{Margaret Drouhard}
\affiliation{%
  \institution{University of Washington}
  \streetaddress{428 Sieg Hall, Box 352315}
  \city{Seattle}
  \state{WA}
  \postcode{98195-2315}
  \country{USA}}
\email{mdrouhard@acm.org}
\authornotemark[1]

\renewcommand{\shortauthors}{Keyes et al.}
\begin{abstract}

The HCI community has worked to expand and improve our consideration of the societal implications of our work and our corresponding responsibilities. Despite this increased engagement, HCI continues to lack an explicitly articulated politic, which we argue re-inscribes and amplifies systemic oppression. In this paper, we set out an explicit political vision of an HCI grounded in emancipatory autonomy\textemdash an \textit{anarchist HCI}, aimed at dismantling all oppressive systems by mandating suspicion of and a reckoning with imbalanced distributions of power. We outline some of the principles and accountability mechanisms that constitute an anarchist HCI. We offer a potential framework for radically reorienting the field towards creating prefigurative counterpower\textemdash systems and spaces that exemplify the world we wish to see, as we go about building the revolution in increment.

\end{abstract}



\maketitle
\input{sections/1_introduction}

\input{sections/2_critical_hci}

\input{sections/3_anarchism}

\input{sections/4_anarchist_hci}

\input{sections/5_world}

\input{sections/6_communities}

\input{sections/7_community}

\input{sections/8_accountability}

\input{sections/9_discussion}

\input{sections/_10_conclusion}

\input{sections/_11_ack}

\bibliographystyle{ACM-Reference-Format}
\balance
\bibliography{ahci}

\end{document}

%% file: sections/1_introduction.tex
\section{Introduction}
\begin{quotation}
\textit{"You are ultimately\textemdash consciously or unconsciously\textemdash  salesmen for a delusive ballet in the ideas of democracy, equal opportunity and free enterprise among people who haven't the possibility of profiting from these."} ~\cite{Illich1968}
\end{quotation}

The last few decades have seen HCI take a turn to examine the societal implications of our work: who is included \cite{Amagno2012, Hankerson2016, Hayes2015, Irani2016}, what values it promotes or embodies~\cite{Shilton2018, Friedman2002, Friedman2017}, and how we respond (or do not) to social shifts~\cite{Light2017}. While this is politically-motivated work, HCI has tended to avoid making our politics \textit{explicit} ~\cite{Lazar2016,Bardzell2015}. 
The result has not been the absence of a politic, but an "implicit neoliberalism"~\cite{Feltwell2018, Dourish2010}.

In this paper, we offer an \textit{explicitly} political HCI\textemdash an \textit{anarchist HCI}\textemdash that reorients the field around
the central principles of autonomy and the justification or elimination of power, with the aim of eliminating oppression. We explore the consequences that such a reorientation would have for our field's norms in relation to the wider systems of the world and the communities in which we engage. Finally, we present some mechanisms to move the field forward and hold ourselves and each other accountable for the impacts of our work.

%% file: sections/2_critical_hci.tex
\section {Critical work in HCI}

\begin{quotation}
\textit{"A critical technical practice will, at least for the foreseeable future, require a split identity -- one foot planted in the craft work of design and the other foot planted in the reflexive work of critique. Successfully spanning these borderlands...will require [work to] support the exploration of alternative work practices that will inevitably seem strange to insiders and outsiders alike. This strangeness will not always be comfortable, but it will be productive nonetheless, both in the esoteric terms of the technical field itself and in the esoteric terms by which we ultimately evaluate a technical field's contribution to society."~\cite{agre1997toward}}
\end{quotation}

As part of the "third wave" of HCI, our field is engaged in an ongoing "turn to the social", described by Rogers as an increasing consideration of the social implications of our work~\cite{rogers2012hci}. The depth of our engagement with this has been limited by our position: HCI straddles both the academy (which frequently shies away from explicitly political positions~\cite{bijker1997bicycles}) and industry (often driven by principles and practices that contraindicate positive social change~\cite{skjolddesign}).

One potential path through these problems is a critical approach to HCI: using theories that feature social, ethical and cultural considerations, along with mechanisms to critique interaction designs and expose their consequences~\cite{rogers2012hci}. Bardzell and Bardzell present such an approach in an overview of humanistic HCI, which they define as "any HCI research or practice that deploys humanistic epistemologies...and methodologies (e.g., critical analysis of designs, processes, and implementations; historical genealogies; conceptual analysis; emancipatory criticism) in service of HCI processes, theories, methods, agenda setting, and practices"~\cite{Bardzell2016}. Referencing Marxist, feminist, postcolonial and psychoanalytic methods of analysis, the Bardzells include within humanistic HCI an "emancipatory HCI", one which is "oriented toward exposing and eradicating one or more forms of bondage and oppression, including structural racism, poverty, sexual repression, colonialism, and other forces/effects of the hegemonic status quo"~\cite{Bardzell2015}. This work draws from components of Shaowen Bardzell's earlier work on feminist HCI, in particular her original conceptualisations
of "pluralism, participation, advocacy, ecology, embodiment, and self-disclosure"~\cite{Bardzell2010}.

Another approach, postcolonial HCI, is exemplified by the works of Lilly Irani~\cite{Irani2010}. Postcolonial HCI considers and deconstructs how colonialism's cultural legacy appears and persists in computing after the termination of colonialism's formal structures~\cite{Philip2010}. In contrast to HCI's traditional focus on "ubiquitous" methods, theories and technologies~\cite{Ahmed2015,Dourish2012}, postcolonial HCI includes critiques of the way actions taken to help the "developing" world often follow the path of capital and private interests. Additionally, it explicitly and actively concerns itself with power relations~\cite{Irani2010}. In contrast to capitalism and ubiquity, postcolonial HCI researchers propose approaches based on social justice \cite{Sun2013}, the centring of indigenous knowledge and users~\cite{Akama2016}, and the development of design paradigms explicitly made, rooted and deployed in local communities, contexts and knowledge~\cite{WinschiersTheophilus2013, Shaw2014}.

Along similar lines, Avle et al. push back strongly against the idea of ``universal'' or ``rational'' design methods, expressing particular concern for how these models may reinscribe colonial relationships~\cite{avle2017methods}. Rosner considers how similar types of design models (hackathons, IDEO, etc.) may limit consideration around design culture by enforcing the idea of the design process as the ``producer of certain kinds of designers: creative, self-sufficient individuals''~\cite{rosner2018critical}. Not only do these conceptions of design challenge popular narratives of the types of artifacts design should produce; they also call into question the way the methods and pedagogy of design have been bounded. Irani's work on IDEO's "design thinking" model notes how it "articulates a racialized understanding of labor, judgment, and the subject and attempts to maintain whiteness at the apex of global hierarchies of labor"~\cite{Irani2018}.
Luiza Prado de O. Martins presents a related critique, feminist speculative design, calling out the risk inherent in claiming an ``apolitical'' position, namely, contributing to the status quo of hierarchies and oppressions, and she cites the particular classism, elitism, and racism that have been propagated through speculative and critical design (SCD). As an alternative, Prado proposes approaching SCD from an intersectional feminist lens in order to explicitly critique and challenge oppressive power structures~\cite{martins2014privilege}.

There are myriad other movements: queer HCI~\cite{Light2011}, post-capitalist HCI~\cite{Feltwell2018}, and anti-oppressive design~\cite{Smyth2014}, each providing their own critiques of HCI's dominant "technochauvinism"~\cite{Broussard2018} and neoliberal ideology. This critical scholarship has sometimes resulted in practical applications
and tools, including Dimond's work on "Hollaback!" (and "feminist HCI for real")~\cite{Dimond2013,Dimond2012tg}, Alsheikh \textit{et al.}'s exploration of postcolonial technology contexts~\cite{Alsheikh2011}, and Fox and Le Dantec's ``Community Historians'' project~\cite{fox2014community}.

Along with many activists on the ground, these researchers and others have applied their respective critical lenses to nurture, support, and hold themselves accountable to the communities in which they live and work. Nevertheless, as with most HCI scholarship, the politics in almost all of these critical works are implicit rather than directly explicated.

Allowing our political stances to remain unspoken has constrained our ability to question and challenge the consequences of the work we put into the world. Further, this silence creates stumbling blocks for critique and accountability mechanisms.
We argue it is imperative that members of the HCI community articulate the respective political foundations of their work, explicitly addressing (a) what state of the world is necessary for the work to realize its intended effect, and (b) what worlds are advanced by its execution. We believe 
such articulations would strengthen the foundation of these and other
critical works, weaving them into a coherent and explicit politic of HCI. Building on Linehan \& Kirman's "anarCHI" paper~\cite{Linehan2014}, along with Asad \textit{et al.}'s "prefigurative design"~\cite{asad2017creating}, we outline our vision of 
one such explicit articulation: an \textit{anarchist HCI}.

%% file: sections/3_anarchism.tex
\section{What is anarchism?}

\begin{quotation}
\textit{``Love labour, hate mastery, and avoid relationship with the government''}~\cite{talmud1}
\end{quotation}
A person confronted with the term ``anarchist'' may find themselves thinking of black-clad, bomb-throwing radicals seeking the destruction of society, an image stemming from the campaigns of ``propaganda of the deed'' in the 1880s~\cite{McLaughlin2016}. But anarchism is far broader than that brief Eurocentric moment in time, constituting a diverse field (not school) of thought aiming to ``root out and eradicate all coercive, hierarchical social relations, and dream up and establish consensual, egalitarian ones in every instance''~\cite{Milstein2010}. Speaking generally, anarchism concerns itself with \textit{power} and \textit{autonomy}. Two core principles of anarchist thought are that autonomy can only be attained through ensuring a consensual basis for power relations, and that human dignity is fundamentally compromised in the absence of autonomy.

So what does this mean in practice? As with any political movement, the answer varies from person to person: anarchist thought covers a broad range of perspectives, philosophies and approaches to autonomy, an appropriate choice given the value \textit{of} autonomy. In this paper we focus on social anarchism, also known as libertarian socialism,\footnote{We would love to discuss other approaches to anarchism, but self-declared anarcho-capitalists ``should be given no more consideration than [other] oxymorons such as a free slave or the living dead''~\cite{Baillargeon2014}, and the only anarcho-primitivist known to not consider literacy a cardinal sin has been in maximum-security prison since 1998~\cite{Kaczynski1995}.} both because of its long theoretical history and the way in which its central principles align with the power-critical and anti-capitalist nature of the HCI works from which we draw. In contrast, \textit{individualist} anarchism (which sees complete individualism without social responsibilities as the ultimate source of dignity) has often been critiqued specifically for failing to engage with power and the "free market", leading ultimately to the resumption of the status quo~\cite{graeber2013democracy, bookchin1995social, nozick}.

One of the major components of social anarchist thought relates to capitalism. As mentioned, anarchism is centred on autonomy and dignity\textemdash and is consequently concerned with the distribution of power. To social anarchists, capitalism's existence fundamentally undermines autonomy and human dignity by embodying unfair power relations~\cite{Baillargeon2014}. Due to its dependence on the commodification, exchange and accrual of goods, capitalism guarantees both inequality between people and a lack of essential resources for some. 
This is not just an incidental or occasional side-effect of a capitalist system\textemdash it is both inevitable and \textit{by design}, because one cannot have a purchaser without an unmet need, or accrual without disparity~\cite{Gelderloos2018}.

At first glance, then, social anarchism appears to simply be Marxism. 
And, indeed, social anarchism has a long history of drawing from Marxist thought (and vice versa) since the First International~\cite{Schmidt2009}. But there is a crucial difference: Marxism assumes the neutrality of the state, and that the oppression that stems from it is a consequence only of the social class that runs it. A "dictatorship of the proletariat", as opposed to one of the landowning classes, is all that is needed to turn it towards the good of humanity~\cite{Milstein2010}.

But anarchists, as discussed, require that every system of power\textemdash not just capital\textemdash justify the ways it compromises individual autonomy for collective autonomy. While some forms of social anarchism
discuss shallow hierarchies as an example of such a justified compromise, social anarchists are unanimous in seeing \textit{the state} as fundamentally dangerous. 
A state is inherently coercive and involuntary~\cite{Wolff1970}, necessarily (by creating a distinction between those vested with power and those giving power up) creates an underclass, and ultimately and inevitably shifts towards centring its own survival over that of any individual citizen under it~\cite{Suissa2018}. Once again, this is not incidental; it is inherent~\cite{Schmidt2013}. This coercion and guarantee of oppression does not change if the state is organised with a purportedly communist economy interwoven, or premised on a different \textit{kind} of dictatorship~\cite{Guerin1970}. 

Instead social anarchists advocate, as Cindy Milstein puts it, ``consensual, egalitarian [social relations] in every instance"~\cite{Milstein2010}, particularly in the form of communalism, self-governing voluntary associations~\cite{Breton2012}, and autonomous\\ zones~\cite{Jeppesen2010}. The product of a person's 
work should be owned by themself~\cite{Lindemann1984}; the tools used to undertake this work should be shared by the community in which it takes place.

Forming these relations, and toppling unjust power structures, is the path towards autonomy and dignity. This work transcends the elimination of state and capital. Imperialism, racism, sexism, ableism, transphobia and other systems of oppression\textemdash systems which underlie and buttress more formal structures\textemdash do not just vanish when the more formalised structures that weaponise them do~\cite{Franks2018}. There is no dignity in a world that lacks capitalism but still features ubiquitous bigotry. For this reason, anarchism has a long historical integration with feminist thought~\cite{Goldman2012,Baillargeon2014, GavinHebert2011}, queer liberation~\cite{Kissack2008, Ackelsberg2013}, anti-racist and anti-imperialist ideologies~\cite{Lasky2011, Van2001} and the intersection thereof~\cite{dupuis2016state, Jeppesen2010, Rogue2012}. Despite its stereotype as a static form of European thought~\cite{Marshall2018}, anarchism has provided part of the theoretical basis for the work of Krishnavarma and Gandhi in India~\cite{FischerTine2015}, the Zapatistas in Mexico~\cite{Lynd2008}, and the political philosophy underpinning the Democratic Federation of Northern Syria (commonly known as Rojava) ~\cite{Leezenberg2017}. In the current era of late-stage capitalism and globalisation, a resurgent anarchism (integrated with other locally-contingent political philosophies across Africa~\cite{Mbah2014}, Asia~\cite{Hwang2016}, and the Americas~\cite{Gomez2008,Lynd2008}) has acted to bring together those to whom existing mechanisms of social order have lost legitimacy~\cite{Kelly2018}, and encouraged the creation of small-scale collectives as well as large-scale political action~\cite{Purkis2013}. As these examples demonstrate, anarchism is easily hybridised; with its focus on autonomy comes a focus on community-appropriate and community-determined approaches to change and governance. In many respects, anarchism is merely "the newest member of a global family that includes numerous historical and present day communal societies and struggles against authority"~\cite{White2004}: even Hobbes saw it as the natural state of human society~\cite{Purcell2016}.

To many, a world lacking states or capitalism sounds utopian\textemdash but anarchists trend towards the pragmatic and applied, \textit{away} from deep theory. The focus on human dignity and autonomy means that the application of anarchist principles to the organisation of day-to-day life cannot wait for some far-off revolution: it must be enacted in the here and now, through prefigurative politics~\cite{Wilson2018, Ince2010}. The revolution comes not on a single day but through the creation of autonomous spaces and forms of organisation, wherever they can take root, both to provide what limited respite they can and because it is through creating these zones\textemdash through invalidating the claim that hierarchies of power are necessary, and through building the \textit{counterpower} of institutions that offer alternatives to non-consensual power relations\textemdash that we go about "forming the structure of the new society within the shell of the old"~\cite{IWW}.

%% file: sections/4_anarchist_hci.tex
\section{Toward an Anarchist HCI}
\begin{quotation}
\textit{"Any significant attempt to decentralize major political and technological institutions...could only happen by overcoming what would surely be powerful resistance to any such policy. It would require something of a revolution."}~\cite{Winner2010}
\end{quotation}

In summary, then, a social anarchist view of the world is that:

\begin{enumerate}
  \item Human dignity is greatest when human autonomy is greatest, and consequently when social relationships are entered into consensually;
  \item Any relationship of power should be held in suspicion and continuously justified, and both a capitalist economic system and a state system of governance fail to justify their excesses;
  \item The solution is the dissolution of both in favour of systems which maximise human autonomy, in a way that centrally recognises all forms of power, including the implicit systems of power such as race, gender, disability and class which underlie formal power structures, and seeks to eliminate them;
  \item This work must be done in a way responsive to local conditions, and in a fashion that is incremental, seeking to build the revolution by creating spaces in the here and now that embody those values.
\end{enumerate}

So what would a field of HCI that is responsive to and built around these principles look like? Primarily, it would be dedicated to building \textit{prefigurative counterpower}: creating constantly-justified spaces that embody autonomous, anti-oppressive values as a means to build the revolution in increment.  This work requires that we, the HCI community, re-examine our core values and radically alter the ways we enact these values in our relationships with each other and the world. 
While we do not wish to prescribe a single path toward this revolution, we elaborate three interconnected threads where we see a need for these relations to be transformed.  

The first and broadest area of scope is our relation with \textit{the world}, defined as the ecological (in the traditional sense) and infrastructural aspects of human existence. Here we would be expected to centre concerns of sustainability, autonomy and control, particularly with regards to how we understand the full range of impacts of our work and the nature of the systems we support. 

Secondly, we will need to reshape our \textit{inter-community} relationships. Over the course of our work, HCI researchers engage with various individuals and communities\textemdash our ``participants.'' An anarchist HCI would approach these sorts of interactions with the intent of allowing appropriate methods and tools to derive from a particular context~\cite{Erete2018}.
We recognize that communities and environments are best understood from within, rather than through a technochauvinistic lens or ``view from nowhere''~\cite{haraway1988situated}. We would rely on methods that are aware of how design and technology have been used to marginalise, and the oppressive nature of the systems we participate in\textemdash methods that actively work to unpick that use and participation~\cite{Smyth2014, CostanzaChock2018}.

The last area, though first in terms of the work we have to do, is \textit{intra-community} relations: how we as HCI researchers and practitioners relate to each other, and the structures we help develop and in which we participate. An anarchist HCI centres power and self-determination: correspondingly, it would necessitate a re-evaluation of inclusivity in our field and of the voices privileged in the processes of design and research. It would require that we demonstratively examine systems of oppression and work to undermine them, including those relating to gender~\cite{Hayes2015}, colonisation~\cite{Irani2016}, racism~\cite{Hankerson2016}, disability~\cite{Shinohara2016}, and class. An anarchist HCI requires an intersectional lens to avoid flattening the experiences of marginalized peoples~\cite{Crenshaw1989, Schlesinger2017}. It would also likely produce new ways of organising, communicating and meeting that are governed by and accessible to the communities concerned.

An anarchist HCI is not merely a conceptual frame. Given its emphasis on prefigurative counterpower, it demands to be brought into being. It demands mechanisms for accountability and justification, adapted to our local context. It demands explicit demonstration that our work is conducted as accomplices rather than overseers and does not act to reinforce systems of power and 
oppression; and that we came ``with empty hands and a desire to unbuild walls''~\cite{Le2015}.

%% file: sections/5_world.tex
\subsection{Global relations}
Given HCI's global reach, a political approach to our work must consider \textit{the world}: the rest of the planet and the (often out of sight) communities and systems that comprise it. Specifically, we need to address how HCI's working practices often presume the universalism of our perspectives, and depend on structures that necessitate the exploitation of labour and resources.

An anarchist HCI is premised on autonomy, not only at the person-to-person level but also of different communities, cultures and contexts. A base requirement of this is an assumption of inherent legitimacy\textemdash that differing ways of being are valid ways of being. One cannot have both autonomy and the exclusive centring of one particular epistemic position. Yet dishearteningly, even within areas of HCI that feature liberatory rhetoric, we find a \textit{universalist} stance. By this we mean that researchers assume their epistemic framings or their experiences within their communal and cultural contexts are "the" human experience. As an example we can take \textit{Gender HCI}~\cite{Beckwith2004}, a subfield concerned with the ways that gendered differences in socialisation make themselves known in technology being more- or less-accessible for differently gendered populations. In theory an anarchist approach to HCI would easily take root here; we care about power and oppression, and differences in technological access which replicate pre-existing inequalities are a quintessential example of that oppression. 

But in practice, Gender HCI is constrained by a particularly narrow vision of gender, and one it treats as universal; with few exceptions~\cite{B2019,Breslin2014,Rode2011,Alsheikh2011}, gender is seen as an essentialist binary in which there are two categories, male and female, with corresponding social and anatomical categories, to which research on gendered differences performed in a Western, academic context is broadly applicable. This approach fundamentally ignores, amongst other things, non-Western models of gender~\cite{Bhaduri2015,Besnier2014,Oyewumi1997}, and the existence of transgender people~\cite{Kannabiran2011,Hayes2015}. Gender HCI research is also frequently undertaken within corporate working environments that assume (or sometimes depend on) top-down action and hierarchy~\cite{Burnett2017, Jose2016}, then assumed to be generalisable to "software" or "gender". In both cases the result is the same\textemdash research premised on universalism that, as a consequence, implicitly delegitimises other ways of being.

An anarchist HCI must shrug off this implicit universalism, not just in relation to gender but in relation to any attribute of a context or individual, in favour of a pluralistic approach in which we interact with other communities on their terms, with an expectation that their members are those best-equipped to define and describe the difficulties being faced. In the case of gender, there are several examples of this approach being done\textemdash in particular Alsheikh \textit{et al}'s work on intimacy in Arab contexts, and Alex Ahmed's work on trans-inclusive interaction design~\cite{Alsheikh2011, Ahmed2017}.
Nonetheless, we have (as other papers note) much progress to make~\cite{Keyes2018, Schlesinger2017}.

HCI's dependence on exploitative global structures can be seen if we examine the predominant cultural conception and practices of
making, which often feature an emancipatory rhetoric of enabling people to autonomously identify their needs and respond to them. Gone are (or will be) the days of mass-produced, industrialised consumer products and tools; instead, every 
home will feature a 3D printer that allows its inhabitants to construct items adapted to their specific use. In theory one might think an anarchist HCI would grab making with both hands as an example of emancipation; after all, don't we have self-determination? A reduction in the inequality of power relations? A reduction in the power of capitalism?

But the problem comes with making's \textit{relation to the rest of the world}: one must ask how emancipatory a technology is, how much autonomy it induces when, for example, it overwhelmingly remains the preserve of those who are already most free. One must also take an ecological and anticolonial bent, as parts of both HCI and anarchism have already done~\cite{Shannon2016, Silberman2014}, and look at the work practices on which making is premised: if a 3D printer is in every house, a truly inhumane amount of copper must have been extracted. And "inhumane" and "extracted" are the right words, because mining is a literally exploitative activity and one that, under capitalism, promotes and perpetuates vast inequalities and injustices. In Chile, which produces a vast amount of the world's copper, the power structures that underpinned mining\textemdash some literally originating in colonial slave labour\textemdash were trivially adapted to solidify Pinochet's military dictatorship~\cite{Frazier2007}. There is no separating out our advocacy and development of making from the costs that making entails\textemdash from the ways that, whatever the emancipatory rhetoric around it, it demands the legitimisation and use of exploitative systems that, beyond their already inhumane day-to-day cost, are so easily twisted into acts of genocide.

This is not specific to making\textemdash indeed, one could argue the \textit{computer} in Human-Computer Interaction means that some amount of exploitation or practical scarcity is inevitable~\cite{Crawford2018}, whatever improvements transpire in ecologically-friendly mining~\cite{Gentina2016}. Nor is our concern solely about ecology: we are simply using copper mining as an example of the global infrastructures that our technology plugs (idiomatically and literally) into. Our point is that our field's existence fuels oppressive systems~\cite{Larkin2013}. This is an inevitable outcome of infrastructures under capitalism, and even absent capitalism, infrastructure enacts control and hegemony~\cite{Bowker2000,Parmiggiani2018}; this is nowhere more apparent than in the infrastructures HCI researchers actively help \textit{build}~\cite{Vaidhyanathan2012,Mager2012}. Consequently from both an anti-oppressive and autonomous perspective, an anarchist HCI is at least highly suspicious of and at most actively opposed to centralised infrastructure. We should avoid making it; we should, wherever possible, avoid participating in it; we should, wherever necessary, actively seek to unmake it.  Winner is right when he says that, absent centralisation, infrastructure and the lopsided benefits that come along with it will be harder to attain, or in some cases impossible~\cite{Winner2010}.

%% file: sections/6_communities.tex
\subsection{Inter-community relations}
Despite the pessimistic note above, we do not mean to suggest that an anarchist HCI inherently opposes \textit{all} infrastructures. Our goal is simply to avoid centralised infrastructure, and challenge systems that accrue power at the expense of human dignity. Given how infrastructures perpetuate their existences and amplify the values encoded within them~\cite{Hacking1996}\textemdash and so are often weaponised for the purposes of hegemony and cultural imperialism~\cite{Vaidhyanathan2012}\textemdash an anarchist HCI requires the constant mapping and justification of infrastructures' power dynamics. Systems that cannot be justified should be supplanted.

In practice this may initially result in a reduction in infrastructure, with associated reductions in the easy transmission of information and goods, but that is largely because of how far (as Winner notes) the pendulum has swung in the direction of centralisation~\cite{Winner2010}. As a prominent example, Ashwin Mathew has tracked how the internet itself is not only centralised but \textit{designed} to be centralised ~\cite{Mathew2016}.

In the long term, there are other ways of running things. Our concern is not \textit{organisation} but who gets to define the terms under which things are organised, and how consensual participation in and departure from systems is: with \textit{autonomy} and \textit{decentralisation}. Rather than an absence of technologies, we are talking about technologies built in a way that centres the communities using them and avoids reserving for some third party the powers to modify, adapt, and repair; about design processes in which the members of that community are treated not as participants but as \textit{accomplices}. In infrastructural terms, that could (to continue the example of the internet) look like distributed replacements, which are already being prototyped~\cite{Sathiaseelan2015, Musiani2014, antoniadis2016}; more generally, it would include open source appropriate technologies (OSATs)\textemdash technologies designed to be low-cost (financially and ecologically), ethically sound, and based around open source software and hardware so that local communities can adapt them to their needs~\cite{Pearce2012}.

But design processes in such an environment have to focus on the needs of the communities as \textit{defined} by those communities: the world contains too many examples of what Meredith Broussard calls  "technochauvinism"~\cite{Broussard2018}\textemdash the deployment of technical solutions against the will or desire of the people subject to them\textemdash for us to be anything but cynical of a top-down approach, even absent an anarchist framework~\cite{Crooks2018, Broussard2018}. Our relationship with local communities should be one in which we \textit{defer}, recognising the centrality of local knowledge in developing local solutions. This consists not only of standalone approaches such as co-design\cite{floyd1989out}, which has been used for large-scale community engagement~\cite{asad2017creating,fox2014community}, but also the adoption of frameworks that recognise pre-existing power relationships and oppression. An example of such a framework is Costanza-Chock's Design Justice (which, interestingly, draws on the example of Zapatismo, a politic that synthesises anarchist principles with indigenous philosophy ~\cite{Costanza2018b}). Design Justice prioritises "projects that challenge the matrix of domination"~\cite{CostanzaChock2018}, focusing on addressing oppression in an intersectional manner (as do many strains of anarchism~\cite{Rogue2012}). Design Justice also aligns with social anarchist principles of autonomy and self-determination due to its focus on local and contextual solutions.

Whether rooted in design theory or anarchist theory, \\localism-based
approaches to design and infrastructure pose their own challenges: aside from efficiencies of scale, issues such as privacy and harassment are potentially harder to handle in infrastructure without centralised oversight ~\cite{Greschbach2012}. And there is always the question of who writes the standards that underpin this infrastructure; how easy it is to reconfigure nodes of, say, a distributed internet, to organise a new network based on new principles. While these (and myriad other) challenges should not be downplayed and must be confronted head-on, distributed and localised infrastructure presents an opportunity to build counterpower by creating autonomous spaces not subject to the centralised control that is inherent to much of modern computing~\cite{Pickerill2016fx}.

%% file: sections/7_community.tex
\subsection{Intra-community relations}

But building this counterpower requires us to engage in prefigurative work: to first organise our own community in alignment with the values of self-determination and consensual, self-organised relations and interactions. We must not only reckon with our contributions to power imbalances in the wider world, but also look unceasingly inward, interrogating how power manifests in our own relationships with each other. It is not possible for us to participate in the making of meaningfully different spaces if we are replicating the same dynamics that have brought us to this point.

We must recognise that our community does not begin any of this work from a ``neutral'' position (as if that were even possible). From a queer, feminist, anticolonial or critical race perspective, our field's norms and methods are inherently laced through with patriarchal, cisnormative, heteronormative beliefs that assume a white and western view of the world. Consider Ahmed \textit{et al.}'s reflective piece on writing for an ACM magazine, in which the ACM, while accepting ad revenue from the U.S. National Security Agency, censored the phrase ``sex worker'' from an accepted piece \textit{about technologies for sex workers}, with the argument that ``ACM is not a political organization''~\cite{Ahmed2018}. Consider the demographics of sex work, and so who, precisely, experiences the most harm from the ACM's deliberate refusal to discuss the existence and rights of sex workers, and sex as a topic. Consider how decisions around language marginalise already-vulnerable people, and that our field already features critiques of its approach to such issues~\cite{Bidwell2016, SsoziMugarura2016}.

More broadly, an examination of our community's priorities, as communicated by the SIGCHI strategic initiatives~\cite{SIGCHI2018b}, raises some difficult questions. For example: if distributing our work beyond our community is a ``core part'' of our values, how do we reconcile that with the ACM charging \$1,700 to make a paper ``Open Access''? With community standards under which making this paper available costs as much as one of the laptops on which it was written? If we care about "local and global HCI", what does it say that even CHI Indonesia publishes its schedule and proceedings in English~\cite{CHIUXID}? What does it say that our annual plan dictates a \textit{minimum} of 3 of the next 5 CHIs be held in Europe and North America~\cite{Terveen2018}? How do we reconcile an initiative aimed at ``supporting and promoting diversity in all its forms'', with spending \$14,000 on inclusion events at our conferences in 2017 \cite{SIGCHI2018}, and the same year, \$24,000 on a communications consultant for "messaging"~\cite{Mentis2017}? A possible explanation for the gap between stated ideals and outcomes can be found in a survey of conference steering committee representatives, where respondents ranked inclusion programming as 5th of 8 possible priorities, estimating an investment of 12 volunteer hours to achieve the steering committees' expectations~\cite{SIGCHI2018c}.

An anarchist HCI would demand a reconfiguration of these failed states, centring access and inclusion.
This is not work that will be completed simply by declaring ourselves anarchists\textemdash see the (often justified) critiques of "manarchism"~\cite{Bottici2017kw,Holyoak2015vx}\textemdash but an anarchist perspective, with its focus on power and dignity as first-order principles of analysis, 
gives us a stronger basis from which to build. This is not work that can justifiably be placed on the shoulders of those who need it. We cannot accept diversity initiatives that take the form, as they do in so much of the academy, of demanding marginalised scholars shoulder the burden of repairing the structural inequities that permeate our institutions ~\cite{Ahmed2012}, or tokenise us in surface-level diversity initiatives that primarily exist for the purpose of public relations~\cite{Ahmed2018b}.

This must be about more than just bodies: it is not diversity if we only accept marginalised people who are stripped of the epistemic models that underpin experiences of being Other, or have the work they draw from those models held to an unequal standard of legitimacy~\cite{Fricker2007}. This must be about plural ways of contributing; plural ways of being present. We would explore different ways of structuring how we gather and conference\textemdash whether expensive and exclusive gatherings of researchers who (speaking practically) have either the employment benefits to attend, contributed a grant-supported paper, or both, act fundamentally as barriers to inclusion and as inducements to inequality. We might look at distributing conferences in their entirety, enabling participation from disparate locations and disparate contexts; we could reorient conferences from closed spaces to open ones, with scholars travelling to talk to the \textit{public} about their work. We could create avenues for publishing that do not operate on the premise that only in English can legitimate science be performed. And if those outside are not interested in participating, when allowed to define the terms of that participation, HCI can hardly argue its work is emancipatory or empowering.

%% file: sections/8_accountability.tex
\section{Accounting for HCI}

\begin{quotation}
\textit{"We must recognize that ethics requires us to risk ourselves precisely at moments of unknowingness, when what forms us diverges from what lies before us, when our willingness to become undone in relation to others constitutes our chance of becoming human."}~\cite{Butler2005}
\end{quotation}

If we want our work to challenge structures of oppression and support human dignity, we are obligated to continually interrogate ways in which our practices and outputs require, perpetuate, or amplify power inequalities. We must work to ensure our technologies actively contribute to (rather than detract from) human autonomy and dignity. Toward these ends, we propose some accountability processes for an anarchist HCI.
 
An anarchist HCI necessarily rejects the premise of a ``neutral technology''~\cite{Gordon2009}. Like many HCI and critical theorists~\cite{feenberg2010between, fallman2011new, verbeek2011moralizing, Winner1980, latour1992missing}, an anarchist HCI seeks to surface the implicit and explicit politics of HCI contributions. While we often expect HCI work to include a researcher stance or reflexive statement in relation to the work, anarchist HCI demands a robust and critical accounting of how we and our work relate to any power structures that oppress people or deprive them of agency. This might manifest as comprehensive, publicly accessible documentation of requirements, intentions, and methods for novel designs\textemdash documentation that proactively demonstrates that the proposed interventions, at a \textit{minimum}, do not reinforce oppressive power structures.

We propose that anarchist HCI should actively contribute to the building of counterpower. Rather than yielding ``responsibility for enabling human flourishing [to] state and corporate actors''~\cite{Light2017}, we argue that it is necessary to actively build systems that undermine such actors, recognising the way that technology metastasises capitalism and the state's worst intentions and vice versa~\cite{Mager2012,Fisher2010, Lyon2003}. Some promising recent work toward these ends include Baumer \& Silberman's proposal not to design~\cite{baumer2011implication}, or Pierce's suggestion that we ``undesign''\textemdash inhibit or foreclose\textemdash particular capabilities of technology~\cite{pierce2012undesigning}.

Given the inevitability of exploitation under capitalism and the state, all work should affirmatively show that it prefigures autonomy and dignity. In other words, that the methods and outputs were driven by the interests and desires of the individuals and communities impacted by the work\textemdash \textit{not} by funders' implicit or explicit expectations. We use "desires", with its implications of subjective, internally-known and validated truth, intentionally: work cannot be undertaken without the active consent and participation of these communities.

One way to incorporate this active consent and participation could be a ``right of participant response'' to research findings and design interventions. In other words, researchers have the ongoing responsibility to provide the research to participants in a comprehensible form. Participants' responses to the work will be considered inherently valid (i.e., they do not require the affirmation of academics), and these responses should be included in whatever form(s) and venue(s) the research is disseminated. Through such an accountability mechanism we might remake HCI to privilege impacted communities. This remaking would contribute to more equitable inter- and intra-community distribution of both participation and the benefits and burdens of design~\cite{CostanzaChock2018}. A related accountability practice could be to alter the peer review process to include community reviewers who can evaluate work intended for publication based on their experiences and comment on the appropriateness of the work's methods, outcomes and consequences. Equitable distribution of benefits and burdens would also necessitate that those community reviewers be compensated fairly for their labour.

These mechanisms would also necessitate that we produce and share knowledge in formats and settings that are appropriate for a given context. It would no longer be considered legitimate for knowledge to be cloistered in the academy, locked behind paywalls or gates, or for academic scholars to be perceived as the sole or primary sources of knowledge production and arbiters of human experience. Rather, in alignment with feminist epistemologies, anarchist HCI would consider knowledge to be situated in particular contexts~\cite{haraway1988situated,harrison2011making}, and would require that the outputs of any knowledge-producing activities or HCI interventions be created in collaboration with and in forms accessible to the communities concerned. Fox and Rosner have put forward one of the forms that dissemination of research might take~\cite{fox2016continuing}, but we would argue that when the community is truly the locus of power, the idea of ``dissemination'' of knowledge may no longer have coherence at all. Instead, communities would determine how to articulate the shared meanings produced during knowledge-making work.

We wish to reiterate that these accountability mechanisms are only part of the work: we have an obligation to institute them but also, as stated, to move beyond them, actively collaborating with communities to break existing systems of injustice and build the world we wish to see. These mechanisms are necessary because they are prefigurative; they are not, in and of themselves, sufficient.

%% file: sections/9_discussion.tex
\section{Discussion}

\begin{quotation}
\textit{"We're setting out from a point of extreme isolation, of extreme weakness. An insurrectional process must be built from the ground up. Nothing appears less likely than an insurrection, but nothing is more necessary."}~\cite{invisible2009}
\end{quotation}

A number of movements within HCI have been working toward situating communities as the locus of power and the arbiters of meaning and value in HCI. We argue that our failure to realize these ideals stems from an inability to reconcile our political rhetoric and critical lenses with the power structures under which we operate. The most promising solution to this dissonance is a rededication of our field towards building prefigurative counterpower.

The justifications, principles, and mechanisms of an anarchist HCI can be used to guide our work and determine whether we are, as a field or individuals, in concordance with our ideals. In outlining these tenets, we do not claim to have created this work out of whole cloth; as discussed, much of this work is \textit{already being done}. What we offer is an articulation of where this work overlaps\textemdash what principles underlie much of it\textemdash and an articulation of processes towards accountability. 
Most importantly, we are drawing a line in the sand, and offering a vision of a present in which nothing is treated as fixed, and by consequence, everything is treated as \textit{possible}.

This is an anarchist vision, but it is not \textit{the} anarchist vision, nor the only political vision HCI could take. In her comments on Linehan \& Kirman's "anarCHI" alt.chi paper, Lilly Irani raised rhetorical questions which we would, slightly rephrased, replicate: what kind of research would you do if you were doing socialist CHI, or libertarian CHI~\cite{Linehan2014}? 
What kinds of research or practice would feature in a \textit{different} anarchist HCI? 

While we have our own biases, believing in a particular political vision centred on autonomy and then \textit{mandating its adoption} would be the height of hypocrisy. In our view, it would violate the anarchist principles we have outlined to dictate either specific implementations or specific practices toward enacting anarchist HCI.  Rather, as we have articulated, we view communities as the rightful decision-makers and loci of power, as they are the entities that can assess their own needs and must live with the consequences of any interventions.  Part of our aim in this work\textemdash and ideally an ongoing process in which the HCI community will engage\textemdash is a deliberate creation of space.  We might envision a rupture in oppressive design paradigms that continually widens as individuals and collectives establish the aims, principles, and practices to architect the worlds they wish to see.
It is worth considering that, despite our proposed reforms to how our community functions, we have no guarantee that it is salvageable: that it can feature equitable power relations while even vaguely resembling its current form. A HCI that centres anti-oppression work and rejects capital and the state will look very different from how it does now\textemdash and may simply not be possible.

But if research into ways of being and organising\textemdash and, for that matter, the ease of drawing on so much of HCI in forming an anarchist vision\textemdash shows us anything, it is that we already have the tools. What normative society often sees as "alternate" or "traditional" ways of configuring our communities and spaces are frequently anarchistic in nature~\cite{Mbah2014, Purcell2017, Ince2010, White2012}; what we see as cutting-edge HCI frequently lends itself to, or in some cases (we would argue) necessitates, an anarchistic approach to structural change. We don't mean to undersell the difficulty or complexity of our task: we are proposing confrontation with vast systems of power. But it is clear at this point that these systems do not work for most of the world: that the only honest alternative to confronting them is acknowledging our field's claims about working for the good of humanity to be a lie.

%% file: sections/_10_conclusion.tex
\section{Conclusion}

\begin{quotation}
\textit{"Remember this: We be many and they be few. They need us more than we need them.\\
\\
Another world is not only possible, she is on her way. On a quiet day, I can hear her breathing."}~\cite{roy2003war}
\end{quotation}

We have presented a vision for a remaking of HCI, one that synthesises theories, methods and fields of study that focus on the dignity, not efficiency, of humanity. With this remaking, designers and technologists are no longer gatekeepers of knowledge or production; we are potential (rather than necessary) collaborators. Our focus is on those marginalised by the way things are, and how we can participate as willing accomplices in the destruction of the perverse machinery that perpetuates this state of affairs. In serving as accomplices, we may find our vocational knowledge and output is valued as secondary to the contributions and perspectives we can offer as individuals, or as members of \textit{our} communities.

We have no certainty as to what happens upon the adoption of an anarchist HCI: what the knock-on effects are of radically remaking our field to prioritise autonomy, self-determination and the justification or reconfiguration of power. But we can only find out by drawing this line in the sand: by asking members of our field to either justify the way things are or join us in changing them. We cannot force you to participate, nor would we want to. All we can do is ask you to decide.
\\
\\
\textit{Which future do you want to help build?}

%% file: sections/_11_ack.tex
\section{Acknowledgements}

We dedicate this paper to those before and around us: to the International Workers of the World; to the Comit\'e de vigilance de Montmartre; to the martyrs of the Columna Durruti; to the residents of KPAM, Fejuve, MAREZ, and Rojava. To Sucheta Goshal; to Noe It\={o}; to Janelle M. Robinson; to Stefon Alexander; to Margret Wander. To those who dared to make spaces where hope could grow, and those who continue to do so.

%% file: anar_paper.bbl

\begin{thebibliography}{152}


\ifx \showCODEN    \undefined \def \showCODEN     #1{\unskip}     \fi
\ifx \showDOI      \undefined \def \showDOI       #1{#1}\fi
\ifx \showISBNx    \undefined \def \showISBNx     #1{\unskip}     \fi
\ifx \showISBNxiii \undefined \def \showISBNxiii  #1{\unskip}     \fi
\ifx \showISSN     \undefined \def \showISSN      #1{\unskip}     \fi
\ifx \showLCCN     \undefined \def \showLCCN      #1{\unskip}     \fi
\ifx \shownote     \undefined \def \shownote      #1{#1}          \fi
\ifx \showarticletitle \undefined \def \showarticletitle #1{#1}   \fi
\ifx \showURL      \undefined \def \showURL       {\relax}        \fi
\providecommand\bibfield[2]{#2}
\providecommand\bibinfo[2]{#2}
\providecommand\natexlab[1]{#1}
\providecommand\showeprint[2][]{arXiv:#2}

\bibitem[\protect\citeauthoryear{Ackelsberg}{Ackelsberg}{2013}]%
        {Ackelsberg2013}
\bibfield{author}{\bibinfo{person}{Martha Ackelsberg}.}
  \bibinfo{year}{2013}\natexlab{}.
\newblock \bibinfo{booktitle}{\emph{Queering Anarchism: Addressing and
  undressing power and desire}}.
\newblock \bibinfo{publisher}{AK Press}.
\newblock


\bibitem[\protect\citeauthoryear{Agre}{Agre}{1997}]%
        {agre1997toward}
\bibfield{author}{\bibinfo{person}{Philip Agre}.}
  \bibinfo{year}{1997}\natexlab{}.
\newblock \showarticletitle{Toward a Critical Technical Practice: Lessons
  Learned in Trying to reform AI}.
\newblock \bibinfo{journal}{\emph{Social Science, Technical Systems and
  Cooperative Work: Beyond the Great Divide. Erlbaum}} (\bibinfo{year}{1997}).
\newblock


\bibitem[\protect\citeauthoryear{Ahmed}{Ahmed}{2018}]%
        {Ahmed2018b}
\bibfield{author}{\bibinfo{person}{Alex Ahmed}.}
  \bibinfo{year}{2018}\natexlab{}.
\newblock \showarticletitle{{Beyond diversity}}.
\newblock \bibinfo{journal}{\emph{Commun. ACM}} \bibinfo{volume}{61},
  \bibinfo{number}{7} (\bibinfo{date}{June} \bibinfo{year}{2018}),
  \bibinfo{pages}{30--32}.
\newblock


\bibitem[\protect\citeauthoryear{Ahmed}{Ahmed}{2017}]%
        {Ahmed2017}
\bibfield{author}{\bibinfo{person}{Alex~A Ahmed}.}
  \bibinfo{year}{2017}\natexlab{}.
\newblock \showarticletitle{{Trans Competent Interaction Design: A Qualitative
  Study on Voice, Identity, and Technology}}.
\newblock \bibinfo{journal}{\emph{Interacting with Computers}}
  \bibinfo{volume}{30}, \bibinfo{number}{1} (\bibinfo{date}{Nov.}
  \bibinfo{year}{2017}), \bibinfo{pages}{53--71}.
\newblock


\bibitem[\protect\citeauthoryear{Ahmed, Almeida, Choi, Pincus, and
  Ireland}{Ahmed et~al\mbox{.}}{2018}]%
        {Ahmed2018}
\bibfield{author}{\bibinfo{person}{Alex~A. Ahmed}, \bibinfo{person}{Teresa
  Almeida}, \bibinfo{person}{Judeth~Oden Choi}, \bibinfo{person}{Jon Pincus},
  {and} \bibinfo{person}{Kelly Ireland}.} \bibinfo{year}{2018}\natexlab{}.
\newblock \showarticletitle{What's at Issue: Sex, Stigma, and Politics in ACM
  Publishing}. In \bibinfo{booktitle}{\emph{Extended Abstracts of the 2018 CHI
  Conference on Human Factors in Computing Systems}}
  \emph{(\bibinfo{series}{CHI EA '18})}. \bibinfo{publisher}{ACM},
  \bibinfo{address}{New York, NY, USA}, Article \bibinfo{articleno}{alt07},
  \bibinfo{numpages}{10}~pages.
\newblock
\showISBNx{978-1-4503-5621-3}
\urldef\tempurl%
\url{https://doi.org/10.1145/3170427.3188400}
\showDOI{\tempurl}


\bibitem[\protect\citeauthoryear{Ahmed}{Ahmed}{2012}]%
        {Ahmed2012}
\bibfield{author}{\bibinfo{person}{Sara Ahmed}.}
  \bibinfo{year}{2012}\natexlab{}.
\newblock \bibinfo{booktitle}{\emph{On being included: Racism and diversity in
  institutional life}}.
\newblock \bibinfo{publisher}{Duke University Press}.
\newblock


\bibitem[\protect\citeauthoryear{Ahmed, Mim, and Jackson}{Ahmed
  et~al\mbox{.}}{2015}]%
        {Ahmed2015}
\bibfield{author}{\bibinfo{person}{Syed~Ishtiaque Ahmed},
  \bibinfo{person}{Nusrat~Jahan Mim}, {and} \bibinfo{person}{Steven~J.
  Jackson}.} \bibinfo{year}{2015}\natexlab{}.
\newblock \showarticletitle{Residual Mobilities: Infrastructural Displacement
  and Post-Colonial Computing in Bangladesh}. In
  \bibinfo{booktitle}{\emph{Proceedings of the 33rd Annual ACM Conference on
  Human Factors in Computing Systems}} \emph{(\bibinfo{series}{CHI '15})}.
  \bibinfo{publisher}{ACM}, \bibinfo{address}{New York, NY, USA},
  \bibinfo{pages}{437--446}.
\newblock
\showISBNx{978-1-4503-3145-6}
\urldef\tempurl%
\url{https://doi.org/10.1145/2702123.2702573}
\showDOI{\tempurl}


\bibitem[\protect\citeauthoryear{Akama, Keen, and West}{Akama
  et~al\mbox{.}}{2016}]%
        {Akama2016}
\bibfield{author}{\bibinfo{person}{Yoko Akama}, \bibinfo{person}{Seth Keen},
  {and} \bibinfo{person}{Peter West}.} \bibinfo{year}{2016}\natexlab{}.
\newblock \showarticletitle{Speculative Design and Heterogeneity in Indigenous
  Nation Building}. In \bibinfo{booktitle}{\emph{Proceedings of the 2016 ACM
  Conference on Designing Interactive Systems}} \emph{(\bibinfo{series}{DIS
  '16})}. \bibinfo{publisher}{ACM}, \bibinfo{address}{New York, NY, USA},
  \bibinfo{pages}{895--899}.
\newblock
\showISBNx{978-1-4503-4031-1}
\urldef\tempurl%
\url{https://doi.org/10.1145/2901790.2901852}
\showDOI{\tempurl}


\bibitem[\protect\citeauthoryear{Alsheikh, Rode, and Lindley}{Alsheikh
  et~al\mbox{.}}{2011}]%
        {Alsheikh2011}
\bibfield{author}{\bibinfo{person}{Tamara Alsheikh},
  \bibinfo{person}{Jennifer~A Rode}, {and} \bibinfo{person}{Si{\^a}n~E
  Lindley}.} \bibinfo{year}{2011}\natexlab{}.
\newblock \showarticletitle{{(Whose) value-sensitive design - a study of
  long-distance relationships in an Arabic cultural context.}}
\newblock \bibinfo{journal}{\emph{ACM Conference on Computer Supported
  Cooperative Work}} (\bibinfo{year}{2011}).
\newblock


\bibitem[\protect\citeauthoryear{Amagno}{Amagno}{2012}]%
        {Amagno2012}
\bibfield{author}{\bibinfo{person}{Gustavo Amagno}.}
  \bibinfo{year}{2012}\natexlab{}.
\newblock \showarticletitle{{The Role of HCI in the Construction of
  Disability}}.
\newblock \bibinfo{journal}{\emph{BCS Human Computer Interaction Conference}}
  (\bibinfo{date}{Aug.} \bibinfo{year}{2012}), \bibinfo{pages}{1--4}.
\newblock


\bibitem[\protect\citeauthoryear{Antoniadis}{Antoniadis}{2016}]%
        {antoniadis2016}
\bibfield{author}{\bibinfo{person}{Panayotis Antoniadis}.}
  \bibinfo{year}{2016}\natexlab{}.
\newblock \showarticletitle{Local networks for local interactions: Four reasons
  why and a way forward}.
\newblock \bibinfo{journal}{\emph{First Monday}} \bibinfo{volume}{21},
  \bibinfo{number}{12} (\bibinfo{year}{2016}).
\newblock


\bibitem[\protect\citeauthoryear{Asad, Le~Dantec, Nielsen, and Diedrick}{Asad
  et~al\mbox{.}}{2017}]%
        {asad2017creating}
\bibfield{author}{\bibinfo{person}{Mariam Asad},
  \bibinfo{person}{Christopher~A. Le~Dantec}, \bibinfo{person}{Becky Nielsen},
  {and} \bibinfo{person}{Kate Diedrick}.} \bibinfo{year}{2017}\natexlab{}.
\newblock \showarticletitle{Creating a Sociotechnical API: Designing City-Scale
  Community Engagement}. In \bibinfo{booktitle}{\emph{Proceedings of the 2017
  CHI Conference on Human Factors in Computing Systems}}
  \emph{(\bibinfo{series}{CHI '17})}. \bibinfo{publisher}{ACM},
  \bibinfo{address}{New York, NY, USA}, \bibinfo{pages}{2295--2306}.
\newblock
\showISBNx{978-1-4503-4655-9}
\urldef\tempurl%
\url{https://doi.org/10.1145/3025453.3025963}
\showDOI{\tempurl}


\bibitem[\protect\citeauthoryear{Avle, Lindtner, and Williams}{Avle
  et~al\mbox{.}}{2017}]%
        {avle2017methods}
\bibfield{author}{\bibinfo{person}{Seyram Avle}, \bibinfo{person}{Silvia
  Lindtner}, {and} \bibinfo{person}{Kaiton Williams}.}
  \bibinfo{year}{2017}\natexlab{}.
\newblock \showarticletitle{{How Methods Make Designers}}. In
  \bibinfo{booktitle}{\emph{Proceedings of the 2017 CHI Conference on Human
  Factors in Computing Systems}}. ACM, \bibinfo{pages}{472--483}.
\newblock


\bibitem[\protect\citeauthoryear{Baillargeon}{Baillargeon}{2014}]%
        {Baillargeon2014}
\bibfield{author}{\bibinfo{person}{Normand Baillargeon}.}
  \bibinfo{year}{2014}\natexlab{}.
\newblock \bibinfo{booktitle}{\emph{{Order Without Power}}}.
\newblock \bibinfo{publisher}{Seven Stories Press}.
\newblock


\bibitem[\protect\citeauthoryear{Bardzell and Bardzell}{Bardzell and
  Bardzell}{2015}]%
        {Bardzell2015}
\bibfield{author}{\bibinfo{person}{Jeffrey Bardzell} {and}
  \bibinfo{person}{Shaowen Bardzell}.} \bibinfo{year}{2015}\natexlab{}.
\newblock \showarticletitle{{Humanistic HCI}}.
\newblock \bibinfo{journal}{\emph{Synthesis Lectures on Human-Centered
  Informatics}} \bibinfo{volume}{8}, \bibinfo{number}{4}
  (\bibinfo{year}{2015}), \bibinfo{pages}{1--185}.
\newblock


\bibitem[\protect\citeauthoryear{Bardzell and Bardzell}{Bardzell and
  Bardzell}{2016}]%
        {Bardzell2016}
\bibfield{author}{\bibinfo{person}{Jeffrey Bardzell} {and}
  \bibinfo{person}{Shaowen Bardzell}.} \bibinfo{year}{2016}\natexlab{}.
\newblock \showarticletitle{Humanistic HCI}.
\newblock \bibinfo{journal}{\emph{interactions}} \bibinfo{volume}{23},
  \bibinfo{number}{2} (\bibinfo{date}{Feb.} \bibinfo{year}{2016}),
  \bibinfo{pages}{20--29}.
\newblock
\showISSN{1072-5520}
\urldef\tempurl%
\url{https://doi.org/10.1145/2888576}
\showDOI{\tempurl}


\bibitem[\protect\citeauthoryear{Bardzell}{Bardzell}{2010}]%
        {Bardzell2010}
\bibfield{author}{\bibinfo{person}{Shaowen Bardzell}.}
  \bibinfo{year}{2010}\natexlab{}.
\newblock \showarticletitle{Feminist HCI: Taking Stock and Outlining an Agenda
  for Design}. In \bibinfo{booktitle}{\emph{Proceedings of the SIGCHI
  Conference on Human Factors in Computing Systems}}
  \emph{(\bibinfo{series}{CHI '10})}. \bibinfo{publisher}{ACM},
  \bibinfo{address}{New York, NY, USA}, \bibinfo{pages}{1301--1310}.
\newblock
\showISBNx{978-1-60558-929-9}
\urldef\tempurl%
\url{https://doi.org/10.1145/1753326.1753521}
\showDOI{\tempurl}


\bibitem[\protect\citeauthoryear{Baumer and Silberman}{Baumer and
  Silberman}{2011}]%
        {baumer2011implication}
\bibfield{author}{\bibinfo{person}{Eric~P.S. Baumer} {and}
  \bibinfo{person}{M.~Six Silberman}.} \bibinfo{year}{2011}\natexlab{}.
\newblock \showarticletitle{When the Implication is Not to Design
  (Technology)}. In \bibinfo{booktitle}{\emph{Proceedings of the SIGCHI
  Conference on Human Factors in Computing Systems}}
  \emph{(\bibinfo{series}{CHI '11})}. \bibinfo{publisher}{ACM},
  \bibinfo{address}{New York, NY, USA}, \bibinfo{pages}{2271--2274}.
\newblock
\showISBNx{978-1-4503-0228-9}
\urldef\tempurl%
\url{https://doi.org/10.1145/1978942.1979275}
\showDOI{\tempurl}


\bibitem[\protect\citeauthoryear{Beckwith and Burnett}{Beckwith and
  Burnett}{2004}]%
        {Beckwith2004}
\bibfield{author}{\bibinfo{person}{Laura Beckwith} {and}
  \bibinfo{person}{Margaret~M Burnett}.} \bibinfo{year}{2004}\natexlab{}.
\newblock \showarticletitle{{Gender - An Important Factor in End-User
  Programming Environments?.}}. In \bibinfo{booktitle}{\emph{2004 IEEE
  Symposium on Visual Languages and Human-Centric Computing (VL/HCC)}}.
  \bibinfo{publisher}{IEEE}, \bibinfo{pages}{107--114}.
\newblock


\bibitem[\protect\citeauthoryear{Besnier and Alexeyeff}{Besnier and
  Alexeyeff}{2014}]%
        {Besnier2014}
\bibfield{author}{\bibinfo{person}{Niko Besnier} {and} \bibinfo{person}{Kalissa
  Alexeyeff}.} \bibinfo{year}{2014}\natexlab{}.
\newblock \bibinfo{booktitle}{\emph{{Gender on the Edge}}}.
\newblock \bibinfo{publisher}{University of Hawaii Press}.
\newblock


\bibitem[\protect\citeauthoryear{Bhaduri and Mukherjee}{Bhaduri and
  Mukherjee}{2015}]%
        {Bhaduri2015}
\bibfield{author}{\bibinfo{person}{Saugata Bhaduri} {and}
  \bibinfo{person}{Indrani Mukherjee}.} \bibinfo{year}{2015}\natexlab{}.
\newblock \bibinfo{booktitle}{\emph{{Transcultural Negotiations of Gender}}}.
\newblock \bibinfo{publisher}{Springer}, \bibinfo{address}{New Delhi}.
\newblock


\bibitem[\protect\citeauthoryear{Bidwell}{Bidwell}{2016}]%
        {Bidwell2016}
\bibfield{author}{\bibinfo{person}{Nicola~J Bidwell}.}
  \bibinfo{year}{2016}\natexlab{}.
\newblock \showarticletitle{{Decolonising HCI and interaction design
  discourse}}.
\newblock \bibinfo{journal}{\emph{XRDS: Crossroads, The ACM Magazine for
  Students}} \bibinfo{volume}{22}, \bibinfo{number}{4} (\bibinfo{date}{June}
  \bibinfo{year}{2016}), \bibinfo{pages}{22--27}.
\newblock


\bibitem[\protect\citeauthoryear{Bijker}{Bijker}{1997}]%
        {bijker1997bicycles}
\bibfield{author}{\bibinfo{person}{Wiebe~E Bijker}.}
  \bibinfo{year}{1997}\natexlab{}.
\newblock \bibinfo{booktitle}{\emph{Of bicycles, bakelites, and bulbs: Toward a
  theory of sociotechnical change}}.
\newblock \bibinfo{publisher}{MIT press}.
\newblock


\bibitem[\protect\citeauthoryear{Bookchin}{Bookchin}{1995}]%
        {bookchin1995social}
\bibfield{author}{\bibinfo{person}{Murray Bookchin}.}
  \bibinfo{year}{1995}\natexlab{}.
\newblock \bibinfo{booktitle}{\emph{Social anarchism or lifestyle anarchism: an
  unbridgeable chasm}}.
\newblock \bibinfo{publisher}{AK Press Edinburgh}.
\newblock


\bibitem[\protect\citeauthoryear{Bottici}{Bottici}{2017}]%
        {Bottici2017kw}
\bibfield{author}{\bibinfo{person}{Chiara Bottici}.}
  \bibinfo{year}{2017}\natexlab{}.
\newblock \showarticletitle{{Bodies in plural: Towards an anarcha-feminist
  manifesto}}.
\newblock \bibinfo{journal}{\emph{Thesis Eleven}} \bibinfo{volume}{142},
  \bibinfo{number}{1} (\bibinfo{date}{Oct.} \bibinfo{year}{2017}),
  \bibinfo{pages}{91--111}.
\newblock


\bibitem[\protect\citeauthoryear{Bowker and Star}{Bowker and Star}{2000}]%
        {Bowker2000}
\bibfield{author}{\bibinfo{person}{Geoffrey~C Bowker} {and}
  \bibinfo{person}{Susan~Leigh Star}.} \bibinfo{year}{2000}\natexlab{}.
\newblock \showarticletitle{{Invisible Mediators of Action: Classification and
  the Ubiquity of Standards}}.
\newblock \bibinfo{journal}{\emph{Mind, Culture, and Activity}}
  \bibinfo{volume}{7}, \bibinfo{number}{1-2} (\bibinfo{date}{May}
  \bibinfo{year}{2000}), \bibinfo{pages}{147--163}.
\newblock


\bibitem[\protect\citeauthoryear{Breslin and Wadhwa}{Breslin and
  Wadhwa}{2014}]%
        {Breslin2014}
\bibfield{author}{\bibinfo{person}{Samantha Breslin} {and}
  \bibinfo{person}{Bimlesh Wadhwa}.} \bibinfo{year}{2014}\natexlab{}.
\newblock \showarticletitle{Exploring Nuanced Gender Perspectives Within the
  HCI Community}. In \bibinfo{booktitle}{\emph{Proceedings of the India HCI
  2014 Conference on Human Computer Interaction}}
  \emph{(\bibinfo{series}{IndiaHCI '14})}. \bibinfo{publisher}{ACM},
  \bibinfo{address}{New York, NY, USA}, Article \bibinfo{articleno}{45},
  \bibinfo{numpages}{10}~pages.
\newblock
\showISBNx{978-1-4503-3218-7}
\urldef\tempurl%
\url{https://doi.org/10.1145/2676702.2676709}
\showDOI{\tempurl}


\bibitem[\protect\citeauthoryear{Breton, Jeppesen, Kruzynski, and
  Sarrasin}{Breton et~al\mbox{.}}{2012}]%
        {Breton2012}
\bibfield{author}{\bibinfo{person}{{\'E}milie Breton}, \bibinfo{person}{Sandra
  Jeppesen}, \bibinfo{person}{Anna Kruzynski}, {and} \bibinfo{person}{Rachel
  Sarrasin}.} \bibinfo{year}{2012}\natexlab{}.
\newblock \showarticletitle{Prefigurative self-governance and
  self-organization: the influence of antiauthoritarian (pro) feminist, radical
  queer and antiracist networks in Quebec}.
\newblock \bibinfo{journal}{\emph{Organize}} (\bibinfo{year}{2012}),
  \bibinfo{pages}{156--173}.
\newblock


\bibitem[\protect\citeauthoryear{Broussard}{Broussard}{2018}]%
        {Broussard2018}
\bibfield{author}{\bibinfo{person}{Meredith Broussard}.}
  \bibinfo{year}{2018}\natexlab{}.
\newblock \bibinfo{booktitle}{\emph{Artificial Unintelligence: How Computers
  Misunderstand the World}}.
\newblock \bibinfo{publisher}{MIT Press}.
\newblock


\bibitem[\protect\citeauthoryear{Burnett, Counts, Lawrence, and Hanson}{Burnett
  et~al\mbox{.}}{2017}]%
        {Burnett2017}
\bibfield{author}{\bibinfo{person}{Margaret Burnett}, \bibinfo{person}{Robin
  Counts}, \bibinfo{person}{Ronette Lawrence}, {and} \bibinfo{person}{Hannah
  Hanson}.} \bibinfo{year}{2017}\natexlab{}.
\newblock \showarticletitle{{Gender HCl and microsoft: Highlights from a
  longitudinal study}}. In \bibinfo{booktitle}{\emph{2017 IEEE Symposium on
  Visual Languages and Human-Centric Computing (VL/HCC)}}.
  \bibinfo{publisher}{IEEE}, \bibinfo{pages}{139--143}.
\newblock


\bibitem[\protect\citeauthoryear{Burnett, Peters, Hill, and Elarief}{Burnett
  et~al\mbox{.}}{2016}]%
        {Jose2016}
\bibfield{author}{\bibinfo{person}{Margaret Burnett}, \bibinfo{person}{Anicia
  Peters}, \bibinfo{person}{Charles Hill}, {and} \bibinfo{person}{Noha
  Elarief}.} \bibinfo{year}{2016}\natexlab{}.
\newblock \showarticletitle{Finding Gender-Inclusiveness Software Issues with
  GenderMag: A Field Investigation}. In \bibinfo{booktitle}{\emph{Proceedings
  of the 2016 CHI Conference on Human Factors in Computing Systems}}
  \emph{(\bibinfo{series}{CHI '16})}. \bibinfo{publisher}{ACM},
  \bibinfo{address}{New York, NY, USA}, \bibinfo{pages}{2586--2598}.
\newblock
\showISBNx{978-1-4503-3362-7}
\urldef\tempurl%
\url{https://doi.org/10.1145/2858036.2858274}
\showDOI{\tempurl}


\bibitem[\protect\citeauthoryear{Butler}{Butler}{2005}]%
        {Butler2005}
\bibfield{author}{\bibinfo{person}{Judith Butler}.}
  \bibinfo{year}{2005}\natexlab{}.
\newblock \bibinfo{booktitle}{\emph{Giving an account of oneself}}.
\newblock \bibinfo{publisher}{Oxford University Press}.
\newblock


\bibitem[\protect\citeauthoryear{CHIuXiD}{CHIuXiD}{2018}]%
        {CHIUXID}
\bibfield{author}{\bibinfo{person}{CHIuXiD}.} \bibinfo{year}{2018}\natexlab{}.
\newblock \bibinfo{title}{Schedule}.
\newblock
\newblock
\urldef\tempurl%
\url{https://web.archive.org/web/20180919072618/http://2018.chiuxid.org/schedule/}
\showURL{%
Retrieved September 18, 2018 from \tempurl}


\bibitem[\protect\citeauthoryear{Costanza-Chock}{Costanza-Chock}{2018a}]%
        {Costanza2018b}
\bibfield{author}{\bibinfo{person}{Sasha Costanza-Chock}.}
  \bibinfo{year}{2018}\natexlab{a}.
\newblock \showarticletitle{Design Justice, AI, and Escape from the Matrix of
  Domination}.
\newblock \bibinfo{journal}{\emph{Journal of Design and Science}}
  (\bibinfo{year}{2018}).
\newblock


\bibitem[\protect\citeauthoryear{Costanza-Chock}{Costanza-Chock}{2018b}]%
        {CostanzaChock2018}
\bibfield{author}{\bibinfo{person}{Sasha Costanza-Chock}.}
  \bibinfo{year}{2018}\natexlab{b}.
\newblock \showarticletitle{{Design Justice: towards an intersectional feminist
  framework for design theory and practice}}.
\newblock \bibinfo{journal}{\emph{Design Research Society}}
  (\bibinfo{date}{March} \bibinfo{year}{2018}), \bibinfo{pages}{1--14}.
\newblock


\bibitem[\protect\citeauthoryear{Crawford and Joler}{Crawford and
  Joler}{2018}]%
        {Crawford2018}
\bibfield{author}{\bibinfo{person}{Kate Crawford} {and} \bibinfo{person}{Vladan
  Joler}.} \bibinfo{year}{2018}\natexlab{}.
\newblock \bibinfo{title}{Anatomy of an AI System}.
\newblock
\newblock
\urldef\tempurl%
\url{https://anatomyof.ai/}
\showURL{%
Retrieved September 18, 2018 from \tempurl}


\bibitem[\protect\citeauthoryear{Crenshaw}{Crenshaw}{1989}]%
        {Crenshaw1989}
\bibfield{author}{\bibinfo{person}{Kimberle Crenshaw}.}
  \bibinfo{year}{1989}\natexlab{}.
\newblock \showarticletitle{Demarginalizing the intersection of race and sex: A
  black feminist critique of antidiscrimination doctrine, feminist theory and
  antiracist politics}.
\newblock \bibinfo{journal}{\emph{University of Chicago Legal Forum}}
  (\bibinfo{year}{1989}), \bibinfo{pages}{139}.
\newblock


\bibitem[\protect\citeauthoryear{Crooks}{Crooks}{0}]%
        {Crooks2018}
\bibfield{author}{\bibinfo{person}{Roderic~N. Crooks}.}
  \bibinfo{year}{0}\natexlab{}.
\newblock \showarticletitle{Times Thirty: Access, Maintenance, and Justice}.
\newblock \bibinfo{journal}{\emph{Science, Technology, \& Human Values}}
  \bibinfo{volume}{0}, \bibinfo{number}{0} (\bibinfo{year}{0}),
  \bibinfo{pages}{0162243918783053}.
\newblock
\urldef\tempurl%
\url{https://doi.org/10.1177/0162243918783053}
\showDOI{\tempurl}
\showeprint{https://doi.org/10.1177/0162243918783053}


\bibitem[\protect\citeauthoryear{Dimond}{Dimond}{2012}]%
        {Dimond2012tg}
\bibfield{author}{\bibinfo{person}{Jill~Patrice Dimond}.}
  \bibinfo{year}{2012}\natexlab{}.
\newblock \emph{\bibinfo{title}{{Feminist HCI for Real}}}.
\newblock \bibinfo{thesistype}{Ph.D. Dissertation}. \bibinfo{school}{Georgia
  Institute of Technology}.
\newblock


\bibitem[\protect\citeauthoryear{Dimond, Dye, Larose, and Bruckman}{Dimond
  et~al\mbox{.}}{2013}]%
        {Dimond2013}
\bibfield{author}{\bibinfo{person}{Jill~P. Dimond},
  \bibinfo{person}{Michaelanne Dye}, \bibinfo{person}{Daphne Larose}, {and}
  \bibinfo{person}{Amy~S. Bruckman}.} \bibinfo{year}{2013}\natexlab{}.
\newblock \showarticletitle{Hollaback!: The Role of Storytelling Online in a
  Social Movement Organization}. In \bibinfo{booktitle}{\emph{Proceedings of
  the 2013 Conference on Computer Supported Cooperative Work}}
  \emph{(\bibinfo{series}{CSCW '13})}. \bibinfo{publisher}{ACM},
  \bibinfo{address}{New York, NY, USA}, \bibinfo{pages}{477--490}.
\newblock
\showISBNx{978-1-4503-1331-5}
\urldef\tempurl%
\url{https://doi.org/10.1145/2441776.2441831}
\showDOI{\tempurl}


\bibitem[\protect\citeauthoryear{Dourish}{Dourish}{2010}]%
        {Dourish2010}
\bibfield{author}{\bibinfo{person}{Paul Dourish}.}
  \bibinfo{year}{2010}\natexlab{}.
\newblock \showarticletitle{HCI and Environmental Sustainability: The Politics
  of Design and the Design of Politics}. In
  \bibinfo{booktitle}{\emph{Proceedings of the 8th ACM Conference on Designing
  Interactive Systems}} \emph{(\bibinfo{series}{DIS '10})}.
  \bibinfo{publisher}{ACM}, \bibinfo{address}{New York, NY, USA},
  \bibinfo{pages}{1--10}.
\newblock
\showISBNx{978-1-4503-0103-9}
\urldef\tempurl%
\url{https://doi.org/10.1145/1858171.1858173}
\showDOI{\tempurl}


\bibitem[\protect\citeauthoryear{Dourish and Mainwaring}{Dourish and
  Mainwaring}{2012}]%
        {Dourish2012}
\bibfield{author}{\bibinfo{person}{Paul Dourish} {and}
  \bibinfo{person}{Scott~D. Mainwaring}.} \bibinfo{year}{2012}\natexlab{}.
\newblock \showarticletitle{Ubicomp's Colonial Impulse}. In
  \bibinfo{booktitle}{\emph{Proceedings of the 2012 ACM Conference on
  Ubiquitous Computing}} \emph{(\bibinfo{series}{UbiComp '12})}.
  \bibinfo{publisher}{ACM}, \bibinfo{address}{New York, NY, USA},
  \bibinfo{pages}{133--142}.
\newblock
\showISBNx{978-1-4503-1224-0}
\urldef\tempurl%
\url{https://doi.org/10.1145/2370216.2370238}
\showDOI{\tempurl}


\bibitem[\protect\citeauthoryear{Dupuis-D{\'e}ri}{Dupuis-D{\'e}ri}{2016}]%
        {dupuis2016state}
\bibfield{author}{\bibinfo{person}{Francis Dupuis-D{\'e}ri}.}
  \bibinfo{year}{2016}\natexlab{}.
\newblock \showarticletitle{Is the State Part of the Matrix of Domination and
  Intersectionality? An Anarchist Inquiry.}
\newblock \bibinfo{journal}{\emph{Anarchist Studies}} \bibinfo{volume}{24},
  \bibinfo{number}{1} (\bibinfo{year}{2016}).
\newblock


\bibitem[\protect\citeauthoryear{Erete, Israni, and Dillahunt}{Erete
  et~al\mbox{.}}{2018}]%
        {Erete2018}
\bibfield{author}{\bibinfo{person}{Sheena Erete}, \bibinfo{person}{Aarti
  Israni}, {and} \bibinfo{person}{Tawanna Dillahunt}.}
  \bibinfo{year}{2018}\natexlab{}.
\newblock \showarticletitle{An Intersectional Approach to Designing in the
  Margins}.
\newblock \bibinfo{journal}{\emph{Interactions}} \bibinfo{volume}{25},
  \bibinfo{number}{3} (\bibinfo{date}{April} \bibinfo{year}{2018}),
  \bibinfo{pages}{66--69}.
\newblock
\showISSN{1072-5520}
\urldef\tempurl%
\url{https://doi.org/10.1145/3194349}
\showDOI{\tempurl}


\bibitem[\protect\citeauthoryear{Fallman}{Fallman}{2011}]%
        {fallman2011new}
\bibfield{author}{\bibinfo{person}{Daniel Fallman}.}
  \bibinfo{year}{2011}\natexlab{}.
\newblock \showarticletitle{The New Good: Exploring the Potential of Philosophy
  of Technology to Contribute to Human-computer Interaction}. In
  \bibinfo{booktitle}{\emph{Proceedings of the SIGCHI Conference on Human
  Factors in Computing Systems}} \emph{(\bibinfo{series}{CHI '11})}.
  \bibinfo{publisher}{ACM}, \bibinfo{address}{New York, NY, USA},
  \bibinfo{pages}{1051--1060}.
\newblock
\showISBNx{978-1-4503-0228-9}
\urldef\tempurl%
\url{https://doi.org/10.1145/1978942.1979099}
\showDOI{\tempurl}


\bibitem[\protect\citeauthoryear{Feenberg}{Feenberg}{2010}]%
        {feenberg2010between}
\bibfield{author}{\bibinfo{person}{Andrew Feenberg}.}
  \bibinfo{year}{2010}\natexlab{}.
\newblock \bibinfo{booktitle}{\emph{{Between Reason and Experience: Essays in
  Technology and Modernity}}}.
\newblock \bibinfo{publisher}{MIT Press}.
\newblock


\bibitem[\protect\citeauthoryear{Feltwell, Lawson, Encinas, Linehan, Kirman,
  Maxwell, Jenkins, and Kuznetsov}{Feltwell et~al\mbox{.}}{2018}]%
        {Feltwell2018}
\bibfield{author}{\bibinfo{person}{Tom Feltwell}, \bibinfo{person}{Shaun
  Lawson}, \bibinfo{person}{Enrique Encinas}, \bibinfo{person}{Conor Linehan},
  \bibinfo{person}{Ben Kirman}, \bibinfo{person}{Deborah Maxwell},
  \bibinfo{person}{Tom Jenkins}, {and} \bibinfo{person}{Stacey Kuznetsov}.}
  \bibinfo{year}{2018}\natexlab{}.
\newblock \showarticletitle{"Grand Visions" for Post-Capitalist Human-Computer
  Interaction}. In \bibinfo{booktitle}{\emph{Extended Abstracts of the 2018 CHI
  Conference on Human Factors in Computing Systems}}
  \emph{(\bibinfo{series}{CHI EA '18})}. \bibinfo{publisher}{ACM},
  \bibinfo{address}{New York, NY, USA}, Article \bibinfo{articleno}{W04},
  \bibinfo{numpages}{8}~pages.
\newblock
\showISBNx{978-1-4503-5621-3}
\urldef\tempurl%
\url{https://doi.org/10.1145/3170427.3170609}
\showDOI{\tempurl}


\bibitem[\protect\citeauthoryear{Fischer-Tin{\'e}}{Fischer-Tin{\'e}}{2015}]%
        {FischerTine2015}
\bibfield{author}{\bibinfo{person}{Harald Fischer-Tin{\'e}}.}
  \bibinfo{year}{2015}\natexlab{}.
\newblock \bibinfo{booktitle}{\emph{{Shyamji Krishnavarma: Sanskrit, Sociology
  and Anti-Imperialism}}}.
\newblock \bibinfo{publisher}{Routledge}.
\newblock


\bibitem[\protect\citeauthoryear{Fisher}{Fisher}{2010}]%
        {Fisher2010}
\bibfield{author}{\bibinfo{person}{Eran Fisher}.}
  \bibinfo{year}{2010}\natexlab{}.
\newblock \bibinfo{booktitle}{\emph{{Media and New Capitalism in the Digital
  Age}}}.
\newblock \bibinfo{publisher}{Springer}, \bibinfo{address}{New York}.
\newblock


\bibitem[\protect\citeauthoryear{Floyd, Mehl, Reisin, Schmidt, and Wolf}{Floyd
  et~al\mbox{.}}{1989}]%
        {floyd1989out}
\bibfield{author}{\bibinfo{person}{Christiane Floyd},
  \bibinfo{person}{Wolf-Michael Mehl}, \bibinfo{person}{Fanny-Michaela Reisin},
  \bibinfo{person}{Gerhard Schmidt}, {and} \bibinfo{person}{Gregor Wolf}.}
  \bibinfo{year}{1989}\natexlab{}.
\newblock \showarticletitle{{Out of Scandinavia: Alternative Approaches to
  Software Design and System Development}}.
\newblock \bibinfo{journal}{\emph{Human-Computer Interaction}}
  \bibinfo{volume}{4}, \bibinfo{number}{4} (\bibinfo{year}{1989}),
  \bibinfo{pages}{253--350}.
\newblock


\bibitem[\protect\citeauthoryear{Fox and Le~Dantec}{Fox and Le~Dantec}{2014}]%
        {fox2014community}
\bibfield{author}{\bibinfo{person}{Sarah Fox} {and}
  \bibinfo{person}{Christopher Le~Dantec}.} \bibinfo{year}{2014}\natexlab{}.
\newblock \showarticletitle{Community Historians: Scaffolding Community
  Engagement Through Culture and Heritage}. In
  \bibinfo{booktitle}{\emph{Proceedings of the 2014 Conference on Designing
  Interactive Systems}} \emph{(\bibinfo{series}{DIS '14})}.
  \bibinfo{publisher}{ACM}, \bibinfo{address}{New York, NY, USA},
  \bibinfo{pages}{785--794}.
\newblock
\showISBNx{978-1-4503-2902-6}
\urldef\tempurl%
\url{https://doi.org/10.1145/2598510.2598563}
\showDOI{\tempurl}


\bibitem[\protect\citeauthoryear{Fox and Rosner}{Fox and Rosner}{2016}]%
        {fox2016continuing}
\bibfield{author}{\bibinfo{person}{Sarah Fox} {and} \bibinfo{person}{Daniela~K
  Rosner}.} \bibinfo{year}{2016}\natexlab{}.
\newblock \showarticletitle{{Continuing the Dialogue: Bringing Research
  Accounts Back into the Field}}. In \bibinfo{booktitle}{\emph{Proceedings of
  the 2016 CHI Conference on Human Factors in Computing Systems}}. ACM,
  \bibinfo{pages}{1426--1430}.
\newblock


\bibitem[\protect\citeauthoryear{Franks, Jun, and Williams}{Franks
  et~al\mbox{.}}{2018}]%
        {Franks2018}
\bibfield{author}{\bibinfo{person}{Benjamin Franks}, \bibinfo{person}{Nathan
  Jun}, {and} \bibinfo{person}{Leonard Williams}.}
  \bibinfo{year}{2018}\natexlab{}.
\newblock \bibinfo{booktitle}{\emph{Anarchism: A Conceptual Approach}}.
\newblock \bibinfo{publisher}{Routledge}.
\newblock


\bibitem[\protect\citeauthoryear{Frazier}{Frazier}{2007}]%
        {Frazier2007}
\bibfield{author}{\bibinfo{person}{Lessie~Jo Frazier}.}
  \bibinfo{year}{2007}\natexlab{}.
\newblock \bibinfo{booktitle}{\emph{Salt in the Sand: Memory, Violence, and the
  Nation-state in Chile, 1890 to the Present}}.
\newblock \bibinfo{publisher}{Duke University Press}.
\newblock


\bibitem[\protect\citeauthoryear{Fricker}{Fricker}{2007}]%
        {Fricker2007}
\bibfield{author}{\bibinfo{person}{Miranda Fricker}.}
  \bibinfo{year}{2007}\natexlab{}.
\newblock \bibinfo{booktitle}{\emph{Epistemic Injustice: Power and the ethics
  of knowing}}.
\newblock \bibinfo{publisher}{Oxford University Press}.
\newblock


\bibitem[\protect\citeauthoryear{Friedman, Hendry, and Borning}{Friedman
  et~al\mbox{.}}{2017}]%
        {Friedman2017}
\bibfield{author}{\bibinfo{person}{Batya Friedman}, \bibinfo{person}{David~G
  Hendry}, {and} \bibinfo{person}{Alan Borning}.}
  \bibinfo{year}{2017}\natexlab{}.
\newblock \showarticletitle{{A Survey of Value Sensitive Design Methods}}.
\newblock \bibinfo{journal}{\emph{Foundations and Trends in
  Human{\textendash}Computer Interaction}} \bibinfo{volume}{11},
  \bibinfo{number}{2} (\bibinfo{year}{2017}), \bibinfo{pages}{63--125}.
\newblock


\bibitem[\protect\citeauthoryear{Friedman, Kahn~Jr, and Borning}{Friedman
  et~al\mbox{.}}{2002}]%
        {Friedman2002}
\bibfield{author}{\bibinfo{person}{Batya Friedman}, \bibinfo{person}{Peter~H
  Kahn~Jr}, {and} \bibinfo{person}{Alan Borning}.}
  \bibinfo{year}{2002}\natexlab{}.
\newblock \bibinfo{booktitle}{\emph{{Value Sensitive Design: Theory and
  Methods}}}.
\newblock \bibinfo{type}{{T}echnical {R}eport}.
\newblock


\bibitem[\protect\citeauthoryear{Gavin-Hebert}{Gavin-Hebert}{2011}]%
        {GavinHebert2011}
\bibfield{author}{\bibinfo{person}{Jane Gavin-Hebert}.}
  \bibinfo{year}{2011}\natexlab{}.
\newblock \emph{\bibinfo{title}{{Plural Desires: Feminist Epistemology as
  Anarchist Praxis}}}.
\newblock \bibinfo{thesistype}{Ph.D. Dissertation}. \bibinfo{school}{Saint
  Mary's University}.
\newblock


\bibitem[\protect\citeauthoryear{Gelderloos}{Gelderloos}{2018}]%
        {Gelderloos2018}
\bibfield{author}{\bibinfo{person}{Peter Gelderloos}.}
  \bibinfo{year}{2018}\natexlab{}.
\newblock \bibinfo{booktitle}{\emph{{Anarchy Works}}}.
\newblock \bibinfo{publisher}{The Anarchist Library}.
\newblock


\bibitem[\protect\citeauthoryear{Gentina and Acevedo}{Gentina and
  Acevedo}{2016}]%
        {Gentina2016}
\bibfield{author}{\bibinfo{person}{Juan Gentina} {and}
  \bibinfo{person}{Fernando Acevedo}.} \bibinfo{year}{2016}\natexlab{}.
\newblock \showarticletitle{{Copper Bioleaching in Chile}}.
\newblock \bibinfo{journal}{\emph{Minerals}} \bibinfo{volume}{6},
  \bibinfo{number}{1} (\bibinfo{date}{March} \bibinfo{year}{2016}),
  \bibinfo{pages}{23--9}.
\newblock


\bibitem[\protect\citeauthoryear{Goldman, de~Cleyre, Dunbar-Ortiz, and
  Freeman}{Goldman et~al\mbox{.}}{2012}]%
        {Goldman2012}
\bibfield{author}{\bibinfo{person}{Emma Goldman}, \bibinfo{person}{Voltairine
  de Cleyre}, \bibinfo{person}{Roxanne Dunbar-Ortiz}, {and} \bibinfo{person}{Jo
  Freeman}.} \bibinfo{year}{2012}\natexlab{}.
\newblock \bibinfo{booktitle}{\emph{Quiet Rumours: An Anarcha-Feminist
  Reader}}.
\newblock \bibinfo{publisher}{AK Press}.
\newblock


\bibitem[\protect\citeauthoryear{G{\'o}mez-Muller}{G{\'o}mez-Muller}{2008}]%
        {Gomez2008}
\bibfield{author}{\bibinfo{person}{Alfredo G{\'o}mez-Muller}.}
  \bibinfo{year}{2008}\natexlab{}.
\newblock \bibinfo{booktitle}{\emph{Anarquismo y anarcosindicalismo en
  Am{\'e}rica Latina.}}
\newblock \bibinfo{publisher}{La Carreta Editores EU}.
\newblock


\bibitem[\protect\citeauthoryear{Gordon}{Gordon}{2009}]%
        {Gordon2009}
\bibfield{author}{\bibinfo{person}{Uri Gordon}.}
  \bibinfo{year}{2009}\natexlab{}.
\newblock \showarticletitle{{Anarchism and the Politics of Technology}}.
\newblock \bibinfo{journal}{\emph{WorkingUSA}} \bibinfo{volume}{12},
  \bibinfo{number}{3} (\bibinfo{date}{Sept.} \bibinfo{year}{2009}),
  \bibinfo{pages}{489--503}.
\newblock


\bibitem[\protect\citeauthoryear{Graeber}{Graeber}{2013}]%
        {graeber2013democracy}
\bibfield{author}{\bibinfo{person}{David Graeber}.}
  \bibinfo{year}{2013}\natexlab{}.
\newblock \bibinfo{booktitle}{\emph{The democracy project: A history, a crisis,
  a movement}}.
\newblock \bibinfo{publisher}{Spiegel \& Grau}.
\newblock


\bibitem[\protect\citeauthoryear{Greschbach, Kreitz, and Buchegger}{Greschbach
  et~al\mbox{.}}{2012}]%
        {Greschbach2012}
\bibfield{author}{\bibinfo{person}{Benjamin Greschbach},
  \bibinfo{person}{Gunnar Kreitz}, {and} \bibinfo{person}{Sonja Buchegger}.}
  \bibinfo{year}{2012}\natexlab{}.
\newblock \showarticletitle{{The devil is in the metadata - New privacy
  challenges in Decentralised Online Social Networks.}}
\newblock \bibinfo{journal}{\emph{PerCom Workshops}} (\bibinfo{year}{2012}),
  \bibinfo{pages}{333--339}.
\newblock


\bibitem[\protect\citeauthoryear{Guerin and Klopper}{Guerin and
  Klopper}{1970}]%
        {Guerin1970}
\bibfield{author}{\bibinfo{person}{Daniel Guerin} {and} \bibinfo{person}{Mary
  Klopper}.} \bibinfo{year}{1970}\natexlab{}.
\newblock \bibinfo{booktitle}{\emph{{Anarchism}}}.
\newblock \bibinfo{publisher}{NYU Press}.
\newblock


\bibitem[\protect\citeauthoryear{Hacking}{Hacking}{1996}]%
        {Hacking1996}
\bibfield{author}{\bibinfo{person}{Ian Hacking}.}
  \bibinfo{year}{1996}\natexlab{}.
\newblock \showarticletitle{{The looping effects of human kinds}}.
\newblock In \bibinfo{booktitle}{\emph{Causal Cognition}}.
  \bibinfo{publisher}{Oxford University Press}, \bibinfo{pages}{351--383}.
\newblock


\bibitem[\protect\citeauthoryear{Hankerson, Rode, Marshall, Booker, El~Mimouni,
  and Walker}{Hankerson et~al\mbox{.}}{2016}]%
        {Hankerson2016}
\bibfield{author}{\bibinfo{person}{David Hankerson},
  \bibinfo{person}{Jennifer~A Rode}, \bibinfo{person}{Andrea~R Marshall},
  \bibinfo{person}{Jennifer Booker}, \bibinfo{person}{Houda El~Mimouni}, {and}
  \bibinfo{person}{Imani Walker}.} \bibinfo{year}{2016}\natexlab{}.
\newblock \showarticletitle{{Does Technology Have Race?}}. In
  \bibinfo{booktitle}{\emph{Conference on Human Factors in Computing Systems}}.
  \bibinfo{publisher}{ACM Press}, \bibinfo{address}{New York, New York, USA},
  \bibinfo{pages}{473--486}.
\newblock


\bibitem[\protect\citeauthoryear{Haraway}{Haraway}{1988}]%
        {haraway1988situated}
\bibfield{author}{\bibinfo{person}{Donna Haraway}.}
  \bibinfo{year}{1988}\natexlab{}.
\newblock \showarticletitle{{Situated Knowledges: The Science Question in
  Feminism and the Privilege of Partial Perspective}}.
\newblock \bibinfo{journal}{\emph{Feminist studies}} \bibinfo{volume}{14},
  \bibinfo{number}{3} (\bibinfo{year}{1988}), \bibinfo{pages}{575--599}.
\newblock


\bibitem[\protect\citeauthoryear{Harrison, Sengers, and Tatar}{Harrison
  et~al\mbox{.}}{2011}]%
        {harrison2011making}
\bibfield{author}{\bibinfo{person}{Steve Harrison}, \bibinfo{person}{Phoebe
  Sengers}, {and} \bibinfo{person}{Deborah Tatar}.}
  \bibinfo{year}{2011}\natexlab{}.
\newblock \showarticletitle{Making epistemological trouble: Third-paradigm
  {HCI} as successor science}.
\newblock \bibinfo{journal}{\emph{Interacting with Computers}}
  \bibinfo{volume}{23}, \bibinfo{number}{5} (\bibinfo{year}{2011}),
  \bibinfo{pages}{385--392}.
\newblock


\bibitem[\protect\citeauthoryear{Hayes and Haimson}{Hayes and Haimson}{2015}]%
        {Hayes2015}
\bibfield{author}{\bibinfo{person}{Gillian~R Hayes} {and}
  \bibinfo{person}{Oliver~L Haimson}.} \bibinfo{year}{2015}\natexlab{}.
\newblock \showarticletitle{{Toward Trans Inclusion in Feminist HCI}}. In
  \bibinfo{booktitle}{\emph{Conference on Human Factors in Computing Systems}}.
  \bibinfo{pages}{8--11}.
\newblock


\bibitem[\protect\citeauthoryear{Holyoak}{Holyoak}{2015}]%
        {Holyoak2015vx}
\bibfield{author}{\bibinfo{person}{Rose~Erin Holyoak}.}
  \bibinfo{year}{2015}\natexlab{}.
\newblock \emph{\bibinfo{title}{{Young Women{\textquoteright}s Gendered
  Subjectivity and Agency in Social Movement Activism Thesis submitted for the
  degree of Doctor of Philosophy at the University of Leicester by Rose Erin
  Holyoak BA (Derby), MA (Nottingham) Department of Sociology University of
  Leicester March 2015}}}.
\newblock \bibinfo{thesistype}{Ph.D. Dissertation}. \bibinfo{school}{University
  of Leicester}.
\newblock


\bibitem[\protect\citeauthoryear{Hwang}{Hwang}{2016}]%
        {Hwang2016}
\bibfield{author}{\bibinfo{person}{Dongyoun Hwang}.}
  \bibinfo{year}{2016}\natexlab{}.
\newblock \bibinfo{booktitle}{\emph{Anarchism in Korea: Independence,
  Transnationalism, and the Question of National Development, 1919-1984}}.
\newblock \bibinfo{publisher}{SUNY Press}.
\newblock


\bibitem[\protect\citeauthoryear{Illich}{Illich}{1968}]%
        {Illich1968}
\bibfield{author}{\bibinfo{person}{Ivan Illich}.}
  \bibinfo{year}{1968}\natexlab{}.
\newblock \showarticletitle{To hell with good intentions}. In
  \bibinfo{booktitle}{\emph{Conference on Inter-American Student Projects}}.
\newblock


\bibitem[\protect\citeauthoryear{Ince}{Ince}{2010}]%
        {Ince2010}
\bibfield{author}{\bibinfo{person}{Anthony James~Elliot Ince}.}
  \bibinfo{year}{2010}\natexlab{}.
\newblock \emph{\bibinfo{title}{{Organising anarchy: spatial strategy
  prefiguration and the politics of everyday life}}}.
\newblock \bibinfo{thesistype}{Ph.D. Dissertation}. \bibinfo{school}{Queen
  Mary, University of London}.
\newblock


\bibitem[\protect\citeauthoryear{invisible}{invisible}{2009}]%
        {invisible2009}
\bibfield{author}{\bibinfo{person}{Comit{\'e} invisible}.}
  \bibinfo{year}{2009}\natexlab{}.
\newblock \bibinfo{booktitle}{\emph{{The coming insurrection}}}.
\newblock \bibinfo{publisher}{Semiotext}.
\newblock


\bibitem[\protect\citeauthoryear{Irani}{Irani}{2018}]%
        {Irani2018}
\bibfield{author}{\bibinfo{person}{Lilly Irani}.}
  \bibinfo{year}{2018}\natexlab{}.
\newblock \showarticletitle{{{\textquotedblleft}Design
  Thinking{\textquotedblright}: Defending Silicon Valley at the Apex of Global
  Labor Hierarchies}}.
\newblock \bibinfo{journal}{\emph{Catalyst: Feminism, Theory, Technoscience}}
  \bibinfo{volume}{4}, \bibinfo{number}{1} (\bibinfo{date}{May}
  \bibinfo{year}{2018}), \bibinfo{pages}{1--19}.
\newblock


\bibitem[\protect\citeauthoryear{Irani, Vertesi, Dourish, Philip, and
  Grinter}{Irani et~al\mbox{.}}{2010}]%
        {Irani2010}
\bibfield{author}{\bibinfo{person}{Lilly Irani}, \bibinfo{person}{Janet
  Vertesi}, \bibinfo{person}{Paul Dourish}, \bibinfo{person}{Kavita Philip},
  {and} \bibinfo{person}{Rebecca~E. Grinter}.} \bibinfo{year}{2010}\natexlab{}.
\newblock \showarticletitle{Postcolonial Computing: A Lens on Design and
  Development}. In \bibinfo{booktitle}{\emph{Proceedings of the SIGCHI
  Conference on Human Factors in Computing Systems}}
  \emph{(\bibinfo{series}{CHI '10})}. \bibinfo{publisher}{ACM},
  \bibinfo{address}{New York, NY, USA}, \bibinfo{pages}{1311--1320}.
\newblock
\showISBNx{978-1-60558-929-9}
\urldef\tempurl%
\url{https://doi.org/10.1145/1753326.1753522}
\showDOI{\tempurl}


\bibitem[\protect\citeauthoryear{Irani and Silberman}{Irani and
  Silberman}{2016}]%
        {Irani2016}
\bibfield{author}{\bibinfo{person}{Lilly~C. Irani} {and}
  \bibinfo{person}{M.~Six Silberman}.} \bibinfo{year}{2016}\natexlab{}.
\newblock \showarticletitle{Stories We Tell About Labor: Turkopticon and the
  Trouble with "Design"}. In \bibinfo{booktitle}{\emph{Proceedings of the 2016
  CHI Conference on Human Factors in Computing Systems}}
  \emph{(\bibinfo{series}{CHI '16})}. \bibinfo{publisher}{ACM},
  \bibinfo{address}{New York, NY, USA}, \bibinfo{pages}{4573--4586}.
\newblock
\showISBNx{978-1-4503-3362-7}
\urldef\tempurl%
\url{https://doi.org/10.1145/2858036.2858592}
\showDOI{\tempurl}


\bibitem[\protect\citeauthoryear{Jeppesen}{Jeppesen}{2010}]%
        {Jeppesen2010}
\bibfield{author}{\bibinfo{person}{Sandra Jeppesen}.}
  \bibinfo{year}{2010}\natexlab{}.
\newblock \showarticletitle{Queer anarchist autonomous zones and publics:
  Direct action vomiting against homonormative consumerism}.
\newblock \bibinfo{journal}{\emph{Sexualities}} \bibinfo{volume}{13},
  \bibinfo{number}{4} (\bibinfo{year}{2010}), \bibinfo{pages}{463--478}.
\newblock


\bibitem[\protect\citeauthoryear{Kaczynski}{Kaczynski}{1995}]%
        {Kaczynski1995}
\bibfield{author}{\bibinfo{person}{Theodore~John Kaczynski}.}
  \bibinfo{year}{1995}\natexlab{}.
\newblock \bibinfo{booktitle}{\emph{Industrial society and its future}}.
\newblock \bibinfo{publisher}{Washington Post}.
\newblock


\bibitem[\protect\citeauthoryear{Kannabiran}{Kannabiran}{2011}]%
        {Kannabiran2011}
\bibfield{author}{\bibinfo{person}{Gopinaath Kannabiran}.}
  \bibinfo{year}{2011}\natexlab{}.
\newblock \showarticletitle{{Themself: Critical Analysis of Gender in
  Facebook}}.
\newblock \bibinfo{journal}{\emph{CHI}} (\bibinfo{year}{2011}),
  \bibinfo{pages}{1--6}.
\newblock


\bibitem[\protect\citeauthoryear{Kelly}{Kelly}{2018}]%
        {Kelly2018}
\bibfield{author}{\bibinfo{person}{Kim Kelly}.}
  \bibinfo{year}{2018}\natexlab{}.
\newblock \bibinfo{title}{Anarchy: What It Is and Why Pop Culture Loves It}.
\newblock
\newblock
\urldef\tempurl%
\url{https://www.teenvogue.com/story/anarchy-explained-what-it-is-why-pop-culture-loves-it}
\showURL{%
Retrieved September 12, 2018 from \tempurl}


\bibitem[\protect\citeauthoryear{Keyes}{Keyes}{2018}]%
        {Keyes2018}
\bibfield{author}{\bibinfo{person}{Os Keyes}.} \bibinfo{year}{2018}\natexlab{}.
\newblock \showarticletitle{The Misgendering Machines: Trans/HCI Implications
  of Automatic Gender Recognition}.
\newblock \bibinfo{journal}{\emph{Proc. ACM Hum.-Comput. Interact.}}
  \bibinfo{volume}{2}, \bibinfo{number}{CSCW}, Article \bibinfo{articleno}{88}
  (\bibinfo{date}{Nov.} \bibinfo{year}{2018}), \bibinfo{numpages}{22}~pages.
\newblock
\showISSN{2573-0142}
\urldef\tempurl%
\url{https://doi.org/10.1145/3274357}
\showDOI{\tempurl}


\bibitem[\protect\citeauthoryear{Kissack}{Kissack}{2008}]%
        {Kissack2008}
\bibfield{author}{\bibinfo{person}{Terence Kissack}.}
  \bibinfo{year}{2008}\natexlab{}.
\newblock \bibinfo{booktitle}{\emph{{Free Comrades: Anarchism and Homosexuality
  in the United States, 0895-0907}}}.
\newblock \bibinfo{publisher}{AK Press}.
\newblock


\bibitem[\protect\citeauthoryear{Larkin}{Larkin}{2013}]%
        {Larkin2013}
\bibfield{author}{\bibinfo{person}{Brian Larkin}.}
  \bibinfo{year}{2013}\natexlab{}.
\newblock \showarticletitle{{The Politics and Poetics of Infrastructure}}.
\newblock \bibinfo{journal}{\emph{Annual Review of Anthropology}}
  \bibinfo{volume}{42}, \bibinfo{number}{1} (\bibinfo{date}{Oct.}
  \bibinfo{year}{2013}), \bibinfo{pages}{327--343}.
\newblock


\bibitem[\protect\citeauthoryear{Lasky}{Lasky}{2011}]%
        {Lasky2011}
\bibfield{author}{\bibinfo{person}{Jackie Lasky}.}
  \bibinfo{year}{2011}\natexlab{}.
\newblock \showarticletitle{Indigenism, anarchism, feminism: an emerging
  framework for exploring post-imperial futures}.
\newblock \bibinfo{journal}{\emph{Affinities: A Journal of Radical Theory,
  Culture, and Action}} (\bibinfo{year}{2011}).
\newblock


\bibitem[\protect\citeauthoryear{Latour}{Latour}{1992}]%
        {latour1992missing}
\bibfield{author}{\bibinfo{person}{Bruno Latour}.}
  \bibinfo{year}{1992}\natexlab{}.
\newblock \showarticletitle{{Where are the Missing Masses? The Sociology of a
  Few Mundane Artifacts}}. In \bibinfo{booktitle}{\emph{Shaping Technology -
  Building Society. Studies in Sociotechnical Change}}. MIT Press,
  \bibinfo{pages}{225--259}.
\newblock


\bibitem[\protect\citeauthoryear{Lazar, Abascal, Barbosa, Barksdale, Friedman,
  Grossklags, Gulliksen, Johnson, McEwan, Mart{\'\i}nez-Normand, Michalk, Tsai,
  van~der Veer, Axelson, Walldius, Whitney, Winckler, Wulf, Churchill, Cranor,
  Davis, Hedge, Hochheiser, Hourcade, Lewis, Nathan, Paterno, Reid, Quesenbery,
  Selker, and Wentz}{Lazar et~al\mbox{.}}{2016}]%
        {Lazar2016}
\bibfield{author}{\bibinfo{person}{Jonathan Lazar}, \bibinfo{person}{Julio
  Abascal}, \bibinfo{person}{Simone Barbosa}, \bibinfo{person}{Jeremy
  Barksdale}, \bibinfo{person}{Batya Friedman}, \bibinfo{person}{Jens
  Grossklags}, \bibinfo{person}{Jan Gulliksen}, \bibinfo{person}{Jeff Johnson},
  \bibinfo{person}{Tom McEwan}, \bibinfo{person}{Lo{\"\i}c
  Mart{\'\i}nez-Normand}, \bibinfo{person}{Wibke Michalk},
  \bibinfo{person}{Janice Tsai}, \bibinfo{person}{Gerrit van~der Veer},
  \bibinfo{person}{Hans Axelson}, \bibinfo{person}{Ake Walldius},
  \bibinfo{person}{Gill Whitney}, \bibinfo{person}{Marco Winckler},
  \bibinfo{person}{Volker Wulf}, \bibinfo{person}{Elizabeth~F Churchill},
  \bibinfo{person}{Lorrie Cranor}, \bibinfo{person}{Janet Davis},
  \bibinfo{person}{Alan Hedge}, \bibinfo{person}{Harry Hochheiser},
  \bibinfo{person}{Juan~Pablo Hourcade}, \bibinfo{person}{Clayton Lewis},
  \bibinfo{person}{Lisa Nathan}, \bibinfo{person}{Fabio Paterno},
  \bibinfo{person}{Blake Reid}, \bibinfo{person}{Whitney Quesenbery},
  \bibinfo{person}{Ted Selker}, {and} \bibinfo{person}{Brian Wentz}.}
  \bibinfo{year}{2016}\natexlab{}.
\newblock \showarticletitle{{Human{\textendash}Computer Interaction and
  International Public Policymaking: A Framework for Understanding and Taking
  Future Actions}}.
\newblock \bibinfo{journal}{\emph{Foundations and Trends in
  Human{\textendash}Computer Interaction}} \bibinfo{volume}{9},
  \bibinfo{number}{2} (\bibinfo{year}{2016}), \bibinfo{pages}{69--149}.
\newblock


\bibitem[\protect\citeauthoryear{Le~Guin}{Le~Guin}{2015}]%
        {Le2015}
\bibfield{author}{\bibinfo{person}{Ursula~K Le~Guin}.}
  \bibinfo{year}{2015}\natexlab{}.
\newblock \bibinfo{booktitle}{\emph{The dispossessed}}.
\newblock \bibinfo{publisher}{Hachette UK}.
\newblock


\bibitem[\protect\citeauthoryear{Leezenberg}{Leezenberg}{2017}]%
        {Leezenberg2017}
\bibfield{author}{\bibinfo{person}{Michiel Leezenberg}.}
  \bibinfo{year}{2017}\natexlab{}.
\newblock \showarticletitle{{The ambiguities of democratic autonomy: the
  Kurdish movement in Turkey and Rojava}}.
\newblock \bibinfo{journal}{\emph{Southeast European and Black Sea Studies}}
  (\bibinfo{date}{Feb.} \bibinfo{year}{2017}), \bibinfo{pages}{0--0}.
\newblock


\bibitem[\protect\citeauthoryear{Light}{Light}{2011}]%
        {Light2011}
\bibfield{author}{\bibinfo{person}{Ann Light}.}
  \bibinfo{year}{2011}\natexlab{}.
\newblock \showarticletitle{{HCI as heterodoxy: Technologies of identity and
  the queering of interaction with computers}}.
\newblock \bibinfo{journal}{\emph{Interacting with Computers}}
  \bibinfo{volume}{23}, \bibinfo{number}{5} (\bibinfo{date}{Sept.}
  \bibinfo{year}{2011}), \bibinfo{pages}{430--438}.
\newblock


\bibitem[\protect\citeauthoryear{Light, Powell, and Shklovski}{Light
  et~al\mbox{.}}{2017}]%
        {Light2017}
\bibfield{author}{\bibinfo{person}{Ann Light}, \bibinfo{person}{Alison Powell},
  {and} \bibinfo{person}{Irina Shklovski}.} \bibinfo{year}{2017}\natexlab{}.
\newblock \showarticletitle{{Design for Existential Crisis in the Anthropocene
  Age}}. In \bibinfo{booktitle}{\emph{ACM International Conference on
  Communities {\&} Technology}}. \bibinfo{publisher}{ACM Press},
  \bibinfo{address}{New York, New York, USA}, \bibinfo{pages}{270--279}.
\newblock


\bibitem[\protect\citeauthoryear{Lindemann}{Lindemann}{1984}]%
        {Lindemann1984}
\bibfield{author}{\bibinfo{person}{Albert~S Lindemann}.}
  \bibinfo{year}{1984}\natexlab{}.
\newblock \bibinfo{booktitle}{\emph{A history of European socialism}}.
\newblock \bibinfo{publisher}{Yale University Press}.
\newblock


\bibitem[\protect\citeauthoryear{Linehan and Kirman}{Linehan and
  Kirman}{2014}]%
        {Linehan2014}
\bibfield{author}{\bibinfo{person}{Conor Linehan} {and} \bibinfo{person}{Ben
  Kirman}.} \bibinfo{year}{2014}\natexlab{}.
\newblock \showarticletitle{{Never mind the bollocks, i wanna be anarCHI}}. In
  \bibinfo{booktitle}{\emph{the extended abstracts of the 32nd annual ACM
  conference}}. \bibinfo{publisher}{ACM Press}, \bibinfo{address}{New York, New
  York, USA}, \bibinfo{pages}{741--748}.
\newblock


\bibitem[\protect\citeauthoryear{Lynd and Grubacic}{Lynd and Grubacic}{2008}]%
        {Lynd2008}
\bibfield{author}{\bibinfo{person}{Staughton Lynd} {and}
  \bibinfo{person}{Andrej Grubacic}.} \bibinfo{year}{2008}\natexlab{}.
\newblock \bibinfo{booktitle}{\emph{{Wobblies and Zapatistas}}}.
\newblock \bibinfo{publisher}{PM Press}.
\newblock


\bibitem[\protect\citeauthoryear{Lyon}{Lyon}{2003}]%
        {Lyon2003}
\bibfield{author}{\bibinfo{person}{David Lyon}.}
  \bibinfo{year}{2003}\natexlab{}.
\newblock \showarticletitle{Surveillance technology and surveillance society}.
\newblock \bibinfo{journal}{\emph{Modernity and technology}}
  (\bibinfo{year}{2003}), \bibinfo{pages}{161--83}.
\newblock


\bibitem[\protect\citeauthoryear{Mager}{Mager}{2012}]%
        {Mager2012}
\bibfield{author}{\bibinfo{person}{Astrid Mager}.}
  \bibinfo{year}{2012}\natexlab{}.
\newblock \showarticletitle{{Algorithmic ideology: How capitalist society
  shapes search engines}}.
\newblock \bibinfo{journal}{\emph{Information, Communication {\&} Society}}
  \bibinfo{volume}{15}, \bibinfo{number}{5} (\bibinfo{date}{June}
  \bibinfo{year}{2012}), \bibinfo{pages}{769--787}.
\newblock


\bibitem[\protect\citeauthoryear{Marshall}{Marshall}{2018}]%
        {Marshall2018}
\bibfield{author}{\bibinfo{person}{Peter Marshall}.}
  \bibinfo{year}{2018}\natexlab{}.
\newblock \showarticletitle{{Demanding the Impossible: A History of
  Anarchism}}.
\newblock  (\bibinfo{date}{Sept.} \bibinfo{year}{2018}),
  \bibinfo{pages}{1--1169}.
\newblock


\bibitem[\protect\citeauthoryear{Martins}{Martins}{2014}]%
        {martins2014privilege}
\bibfield{author}{\bibinfo{person}{Luiza Prado de~O Martins}.}
  \bibinfo{year}{2014}\natexlab{}.
\newblock \showarticletitle{Privilege and oppression: Towards a feminist
  speculative design}.
\newblock \bibinfo{journal}{\emph{Proceedings of DRS}} (\bibinfo{year}{2014}),
  \bibinfo{pages}{980--990}.
\newblock


\bibitem[\protect\citeauthoryear{Mathew}{Mathew}{2016}]%
        {Mathew2016}
\bibfield{author}{\bibinfo{person}{Ashwin~J Mathew}.}
  \bibinfo{year}{2016}\natexlab{}.
\newblock \showarticletitle{{The myth of the decentralised internet}}.
\newblock \bibinfo{journal}{\emph{Internet Policy Review}}
  (\bibinfo{date}{Sept.} \bibinfo{year}{2016}).
\newblock


\bibitem[\protect\citeauthoryear{Mbah and Bufe}{Mbah and Bufe}{2014}]%
        {Mbah2014}
\bibfield{author}{\bibinfo{person}{Sam Mbah} {and} \bibinfo{person}{Chaz
  Bufe}.} \bibinfo{year}{2014}\natexlab{}.
\newblock \bibinfo{booktitle}{\emph{{African Anarchism}}}.
\newblock \bibinfo{publisher}{See Sharp Press}.
\newblock


\bibitem[\protect\citeauthoryear{McLaughlin}{McLaughlin}{2016}]%
        {McLaughlin2016}
\bibfield{author}{\bibinfo{person}{Paul McLaughlin}.}
  \bibinfo{year}{2016}\natexlab{}.
\newblock \bibinfo{booktitle}{\emph{{Anarchism and Authority}}}.
\newblock \bibinfo{publisher}{Routledge}.
\newblock


\bibitem[\protect\citeauthoryear{Mentis}{Mentis}{2017}]%
        {Mentis2017}
\bibfield{author}{\bibinfo{person}{Helena Mentis}.}
  \bibinfo{year}{2017}\natexlab{}.
\newblock \bibinfo{title}{Executive Committee Meeting Notes}.
\newblock
\newblock
\urldef\tempurl%
\url{https://web.archive.org/web/20180919075058-/https://sigchi.org/wp-content/uploads/2017/03/SIGCHI-EC-meeting-notes-201507-presentation.pdf}
\showURL{%
Retrieved September 18, 2018 from \tempurl}


\bibitem[\protect\citeauthoryear{Milstein}{Milstein}{2010}]%
        {Milstein2010}
\bibfield{author}{\bibinfo{person}{Cindy Milstein}.}
  \bibinfo{year}{2010}\natexlab{}.
\newblock \bibinfo{booktitle}{\emph{{Anarchism and Its Aspirations}}}.
\newblock \bibinfo{publisher}{AK Press}.
\newblock


\bibitem[\protect\citeauthoryear{Musiani}{Musiani}{2014}]%
        {Musiani2014}
\bibfield{author}{\bibinfo{person}{F Musiani}.}
  \bibinfo{year}{2014}\natexlab{}.
\newblock \showarticletitle{{Decentralised Internet governance: The case of a
  'peer-to-peer cloud'}}.
\newblock \bibinfo{journal}{\emph{Internet Policy Review}}
  (\bibinfo{year}{2014}).
\newblock


\bibitem[\protect\citeauthoryear{Nozick}{Nozick}{1974}]%
        {nozick}
\bibfield{author}{\bibinfo{person}{Robert Nozick}.}
  \bibinfo{year}{1974}\natexlab{}.
\newblock \bibinfo{booktitle}{\emph{Anarchy, State and Utopia}}.
\newblock \bibinfo{publisher}{Basic Books}.
\newblock


\bibitem[\protect\citeauthoryear{Oy{\v e}w{\`u}m{\'\i}}{Oy{\v
  e}w{\`u}m{\'\i}}{1997}]%
        {Oyewumi1997}
\bibfield{author}{\bibinfo{person}{Oy{\`e}r{\'o}nk\d{\'{e}} Oy{\v
  e}w{\`u}m{\'\i}}.} \bibinfo{year}{1997}\natexlab{}.
\newblock \bibinfo{booktitle}{\emph{{The Invention of Women}}}.
\newblock \bibinfo{publisher}{U of Minnesota Press}.
\newblock


\bibitem[\protect\citeauthoryear{Parmiggiani and Karasti}{Parmiggiani and
  Karasti}{2018}]%
        {Parmiggiani2018}
\bibfield{author}{\bibinfo{person}{Elena Parmiggiani} {and}
  \bibinfo{person}{Helena Karasti}.} \bibinfo{year}{2018}\natexlab{}.
\newblock \showarticletitle{Surfacing the Arctic: Politics of Participation in
  Infrastructuring}. In \bibinfo{booktitle}{\emph{Proceedings of the 15th
  Participatory Design Conference: Short Papers, Situated Actions, Workshops
  and Tutorial - Volume 2}} \emph{(\bibinfo{series}{PDC '18})}.
  \bibinfo{publisher}{ACM}, \bibinfo{address}{New York, NY, USA}, Article
  \bibinfo{articleno}{7}, \bibinfo{numpages}{5}~pages.
\newblock
\showISBNx{978-1-4503-5574-2}
\urldef\tempurl%
\url{https://doi.org/10.1145/3210604.3210625}
\showDOI{\tempurl}


\bibitem[\protect\citeauthoryear{Pearce}{Pearce}{2012}]%
        {Pearce2012}
\bibfield{author}{\bibinfo{person}{Joshua~M Pearce}.}
  \bibinfo{year}{2012}\natexlab{}.
\newblock \showarticletitle{{The case for open source appropriate technology}}.
\newblock \bibinfo{journal}{\emph{Environment, Development and Sustainability}}
  \bibinfo{volume}{14}, \bibinfo{number}{3} (\bibinfo{date}{Jan.}
  \bibinfo{year}{2012}), \bibinfo{pages}{425--431}.
\newblock


\bibitem[\protect\citeauthoryear{Philip, Irani, and Dourish}{Philip
  et~al\mbox{.}}{2010}]%
        {Philip2010}
\bibfield{author}{\bibinfo{person}{Kavita Philip}, \bibinfo{person}{Lilly
  Irani}, {and} \bibinfo{person}{Paul Dourish}.}
  \bibinfo{year}{2010}\natexlab{}.
\newblock \showarticletitle{{Postcolonial Computing}}.
\newblock \bibinfo{journal}{\emph{Science, Technology, {\&} Human Values}}
  \bibinfo{volume}{37}, \bibinfo{number}{1} (\bibinfo{date}{Nov.}
  \bibinfo{year}{2010}), \bibinfo{pages}{3--29}.
\newblock


\bibitem[\protect\citeauthoryear{Pickerill and Chatterton}{Pickerill and
  Chatterton}{2016}]%
        {Pickerill2016fx}
\bibfield{author}{\bibinfo{person}{Jenny Pickerill} {and} \bibinfo{person}{Paul
  Chatterton}.} \bibinfo{year}{2016}\natexlab{}.
\newblock \showarticletitle{{Notes towards autonomous geographies: creation,
  resistance and self-management as survival tactics}}.
\newblock \bibinfo{journal}{\emph{Progress in Human Geography}}
  \bibinfo{volume}{30}, \bibinfo{number}{6} (\bibinfo{date}{July}
  \bibinfo{year}{2016}), \bibinfo{pages}{730--746}.
\newblock


\bibitem[\protect\citeauthoryear{Pierce}{Pierce}{2012}]%
        {pierce2012undesigning}
\bibfield{author}{\bibinfo{person}{James Pierce}.}
  \bibinfo{year}{2012}\natexlab{}.
\newblock \showarticletitle{Undesigning Technology: Considering the Negation of
  Design by Design}. In \bibinfo{booktitle}{\emph{Proceedings of the SIGCHI
  Conference on Human Factors in Computing Systems}}
  \emph{(\bibinfo{series}{CHI '12})}. \bibinfo{publisher}{ACM},
  \bibinfo{address}{New York, NY, USA}, \bibinfo{pages}{957--966}.
\newblock
\showISBNx{978-1-4503-1015-4}
\urldef\tempurl%
\url{https://doi.org/10.1145/2207676.2208540}
\showDOI{\tempurl}


\bibitem[\protect\citeauthoryear{Purcell}{Purcell}{2016}]%
        {Purcell2016}
\bibfield{author}{\bibinfo{person}{Mark Purcell}.}
  \bibinfo{year}{2016}\natexlab{}.
\newblock \showarticletitle{{For democracy: Planning and publics without the
  state}}.
\newblock \bibinfo{journal}{\emph{Planning Theory}} \bibinfo{volume}{15},
  \bibinfo{number}{4} (\bibinfo{date}{Oct.} \bibinfo{year}{2016}),
  \bibinfo{pages}{386--401}.
\newblock


\bibitem[\protect\citeauthoryear{Purcell}{Purcell}{2017}]%
        {Purcell2017}
\bibfield{author}{\bibinfo{person}{Mark Purcell}.}
  \bibinfo{year}{2017}\natexlab{}.
\newblock \showarticletitle{{Our Own Power to Act}}.
\newblock \bibinfo{journal}{\emph{Planning Theory {\&} Practice}}
  (\bibinfo{date}{Oct.} \bibinfo{year}{2017}), \bibinfo{pages}{1--5}.
\newblock


\bibitem[\protect\citeauthoryear{Purkis and Bowen}{Purkis and Bowen}{2013}]%
        {Purkis2013}
\bibfield{author}{\bibinfo{person}{Jonathan Purkis} {and}
  \bibinfo{person}{James Bowen}.} \bibinfo{year}{2013}\natexlab{}.
\newblock \bibinfo{booktitle}{\emph{{Changing Anarchism}}}.
\newblock \bibinfo{publisher}{Oxford University Press}.
\newblock


\bibitem[\protect\citeauthoryear{Rode}{Rode}{2011}]%
        {Rode2011}
\bibfield{author}{\bibinfo{person}{Jennifer~A. Rode}.}
  \bibinfo{year}{2011}\natexlab{}.
\newblock \showarticletitle{A Theoretical Agenda for Feminist HCI}.
\newblock \bibinfo{journal}{\emph{Interact. Comput.}} \bibinfo{volume}{23},
  \bibinfo{number}{5} (\bibinfo{date}{Sept.} \bibinfo{year}{2011}),
  \bibinfo{pages}{393--400}.
\newblock
\showISSN{0953-5438}
\urldef\tempurl%
\url{https://doi.org/10.1016/j.intcom.2011.04.005}
\showDOI{\tempurl}


\bibitem[\protect\citeauthoryear{Rogers}{Rogers}{2012}]%
        {rogers2012hci}
\bibfield{author}{\bibinfo{person}{Yvonne Rogers}.}
  \bibinfo{year}{2012}\natexlab{}.
\newblock \showarticletitle{{HCI Theory: Classical, Modern, and Contemporary}}.
\newblock \bibinfo{journal}{\emph{Synthesis Lectures on Human-Centered
  Informatics}} \bibinfo{volume}{5}, \bibinfo{number}{2}
  (\bibinfo{year}{2012}), \bibinfo{pages}{1--129}.
\newblock
\urldef\tempurl%
\url{https://doi.org/10.2200/S00418ED1V01Y201205HCI014}
\showDOI{\tempurl}
\showeprint{http://dx.doi.org/10.2200/S00418ED1V01Y201205HCI014}


\bibitem[\protect\citeauthoryear{Rogue and Volcano}{Rogue and Volcano}{2012}]%
        {Rogue2012}
\bibfield{author}{\bibinfo{person}{J Rogue} {and} \bibinfo{person}{Abbey
  Volcano}.} \bibinfo{year}{2012}\natexlab{}.
\newblock \showarticletitle{Insurrection at the intersections: Feminism,
  intersectionality, and anarchism}.
\newblock In \bibinfo{booktitle}{\emph{Quiet rumors: An anarcha-feminist
  reader}}, \bibfield{editor}{\bibinfo{person}{Jan Fagerberg},
  \bibinfo{person}{David~C. Mowery}, {and} \bibinfo{person}{Richard~R. Nelson}}
  (Eds.). \bibinfo{publisher}{AK Press Oakland, CA}, \bibinfo{pages}{43--46}.
\newblock


\bibitem[\protect\citeauthoryear{Rosner}{Rosner}{2018}]%
        {rosner2018critical}
\bibfield{author}{\bibinfo{person}{Daniela~K Rosner}.}
  \bibinfo{year}{2018}\natexlab{}.
\newblock \bibinfo{booktitle}{\emph{Critical Fabulations: Reworking the Methods
  and Margins of Design}}.
\newblock \bibinfo{publisher}{MIT Press}.
\newblock


\bibitem[\protect\citeauthoryear{Roy}{Roy}{2003}]%
        {roy2003war}
\bibfield{author}{\bibinfo{person}{Arundhati Roy}.}
  \bibinfo{year}{2003}\natexlab{}.
\newblock \bibinfo{booktitle}{\emph{War Talk}}.
\newblock \bibinfo{publisher}{South End Press}.
\newblock


\bibitem[\protect\citeauthoryear{Sathiaseelan, Wang, Aucinas, Tyson, and
  Crowcroft}{Sathiaseelan et~al\mbox{.}}{2015}]%
        {Sathiaseelan2015}
\bibfield{author}{\bibinfo{person}{Arjuna Sathiaseelan}, \bibinfo{person}{Liang
  Wang}, \bibinfo{person}{Andrius Aucinas}, \bibinfo{person}{Gareth Tyson},
  {and} \bibinfo{person}{Jon Crowcroft}.} \bibinfo{year}{2015}\natexlab{}.
\newblock \showarticletitle{SCANDEX: Service Centric Networking for Challenged
  Decentralised Networks}. In \bibinfo{booktitle}{\emph{Proceedings of the 2015
  Workshop on Do-it-yourself Networking: An Interdisciplinary Approach}}
  \emph{(\bibinfo{series}{DIYNetworking '15})}. \bibinfo{publisher}{ACM},
  \bibinfo{address}{New York, NY, USA}, \bibinfo{pages}{15--20}.
\newblock
\showISBNx{978-1-4503-3503-4}
\urldef\tempurl%
\url{https://doi.org/10.1145/2753488.2753490}
\showDOI{\tempurl}


\bibitem[\protect\citeauthoryear{Schlesinger, Edwards, and Grinter}{Schlesinger
  et~al\mbox{.}}{2017}]%
        {Schlesinger2017}
\bibfield{author}{\bibinfo{person}{Ari Schlesinger}, \bibinfo{person}{W.~Keith
  Edwards}, {and} \bibinfo{person}{Rebecca~E. Grinter}.}
  \bibinfo{year}{2017}\natexlab{}.
\newblock \showarticletitle{Intersectional HCI: Engaging Identity Through
  Gender, Race, and Class}. In \bibinfo{booktitle}{\emph{Proceedings of the
  2017 CHI Conference on Human Factors in Computing Systems}}
  \emph{(\bibinfo{series}{CHI '17})}. \bibinfo{publisher}{ACM},
  \bibinfo{address}{New York, NY, USA}, \bibinfo{pages}{5412--5427}.
\newblock
\showISBNx{978-1-4503-4655-9}
\urldef\tempurl%
\url{https://doi.org/10.1145/3025453.3025766}
\showDOI{\tempurl}


\bibitem[\protect\citeauthoryear{Schmidt}{Schmidt}{2009}]%
        {Schmidt2009}
\bibfield{author}{\bibinfo{person}{Michael Schmidt}.}
  \bibinfo{year}{2009}\natexlab{}.
\newblock \showarticletitle{Black Flame: The Revolutionary Class Politics of
  Anarchism and Syndicalism (Counter-Power, Volume 1)}.
\newblock  (\bibinfo{year}{2009}).
\newblock


\bibitem[\protect\citeauthoryear{Schmidt}{Schmidt}{2013}]%
        {Schmidt2013}
\bibfield{author}{\bibinfo{person}{Michael Schmidt}.}
  \bibinfo{year}{2013}\natexlab{}.
\newblock \bibinfo{booktitle}{\emph{{Cartography of Revolutionary Anarchism}}}.
\newblock \bibinfo{publisher}{AK Press}.
\newblock


\bibitem[\protect\citeauthoryear{Shannon}{Shannon}{2016}]%
        {Shannon2016}
\bibfield{author}{\bibinfo{person}{Deric Shannon}.}
  \bibinfo{year}{2016}\natexlab{}.
\newblock \showarticletitle{{Intersectionality, ecology, food: Conflict
  theory{\textquoteright}s missing lens}}.
\newblock In \bibinfo{booktitle}{\emph{Emergent Possibilities for Global
  Sustainability: Intersections of Race, Class and Gender}}.
  \bibinfo{publisher}{Routledge}, \bibinfo{pages}{39--49}.
\newblock


\bibitem[\protect\citeauthoryear{Shaw, Brereton, and Roe}{Shaw
  et~al\mbox{.}}{2014}]%
        {Shaw2014}
\bibfield{author}{\bibinfo{person}{Grace Shaw}, \bibinfo{person}{Margot
  Brereton}, {and} \bibinfo{person}{Paul Roe}.}
  \bibinfo{year}{2014}\natexlab{}.
\newblock \showarticletitle{Mobile Phone Use in Australian Indigenous
  Communities: Future Pathways for HCI4D}. In
  \bibinfo{booktitle}{\emph{Proceedings of the 26th Australian Computer-Human
  Interaction Conference on Designing Futures: The Future of Design}}
  \emph{(\bibinfo{series}{OzCHI '14})}. \bibinfo{publisher}{ACM},
  \bibinfo{address}{New York, NY, USA}, \bibinfo{pages}{480--483}.
\newblock
\showISBNx{978-1-4503-0653-9}
\urldef\tempurl%
\url{https://doi.org/10.1145/2686612.2686688}
\showDOI{\tempurl}


\bibitem[\protect\citeauthoryear{Shemaya}{Shemaya}{189}]%
        {talmud1}
\bibfield{author}{\bibinfo{person}{Shemaya}.} \bibinfo{year}{189}\natexlab{}.
\newblock \showarticletitle{Pirkei Avot}.
\newblock In \bibinfo{booktitle}{\emph{Mishnah}},
  \bibfield{editor}{\bibinfo{person}{Judah HaNasi}} (Ed.). Chapter~1.
\newblock


\bibitem[\protect\citeauthoryear{Shilton}{Shilton}{2018}]%
        {Shilton2018}
\bibfield{author}{\bibinfo{person}{Katie Shilton}.}
  \bibinfo{year}{2018}\natexlab{}.
\newblock \showarticletitle{{Values and Ethics in Human-Computer Interaction}}.
\newblock \bibinfo{journal}{\emph{Foundations and Trends in
  Human{\textendash}Computer Interaction}} \bibinfo{volume}{12},
  \bibinfo{number}{2} (\bibinfo{year}{2018}), \bibinfo{pages}{107--171}.
\newblock


\bibitem[\protect\citeauthoryear{Shinohara, Bennett, and Wobbrock}{Shinohara
  et~al\mbox{.}}{2016}]%
        {Shinohara2016}
\bibfield{author}{\bibinfo{person}{Kristen Shinohara},
  \bibinfo{person}{Cynthia~L. Bennett}, {and} \bibinfo{person}{Jacob~O.
  Wobbrock}.} \bibinfo{year}{2016}\natexlab{}.
\newblock \showarticletitle{How Designing for People With and Without
  Disabilities Shapes Student Design Thinking}. In
  \bibinfo{booktitle}{\emph{Proceedings of the 18th International ACM SIGACCESS
  Conference on Computers and Accessibility}} \emph{(\bibinfo{series}{ASSETS
  '16})}. \bibinfo{publisher}{ACM}, \bibinfo{address}{New York, NY, USA},
  \bibinfo{pages}{229--237}.
\newblock
\showISBNx{978-1-4503-4124-0}
\urldef\tempurl%
\url{https://doi.org/10.1145/2982142.2982158}
\showDOI{\tempurl}


\bibitem[\protect\citeauthoryear{SIGCHI}{SIGCHI}{2018a}]%
        {SIGCHI2018}
\bibfield{author}{\bibinfo{person}{SIGCHI}.} \bibinfo{year}{2018}\natexlab{a}.
\newblock \bibinfo{title}{Internationalisation, Diversity and Inclusion events
  at Sponsored Conferences}.
\newblock
\newblock
\urldef\tempurl%
\url{https://web.archive.org/web/20180919195612/https://sigchi.org/2017/06/internationalisation\%2Ddiversity\%2Dand\%2Dinclusion\%2Devents\%2Dat\%2Dsponsored-conferences/}
\showURL{%
Retrieved September 18, 2018 from \tempurl}


\bibitem[\protect\citeauthoryear{SIGCHI}{SIGCHI}{2018b}]%
        {SIGCHI2018c}
\bibfield{author}{\bibinfo{person}{SIGCHI}.} \bibinfo{year}{2018}\natexlab{b}.
\newblock \bibinfo{title}{SIGCHI Conferences Report}.
\newblock
\newblock
\urldef\tempurl%
\url{https://web.archive.org/web/20180920073001/'-'https://sigchi.org/wp-content/uploads/2018/05/SIGCHI-Conferences-Report-for-Members-May-2018.pdf}
\showURL{%
Retrieved September 19, 2018 from \tempurl}


\bibitem[\protect\citeauthoryear{SIGCHI}{SIGCHI}{2018c}]%
        {SIGCHI2018b}
\bibfield{author}{\bibinfo{person}{SIGCHI}.} \bibinfo{year}{2018}\natexlab{c}.
\newblock \bibinfo{title}{SIGCHI EC Values and Strategic Initiatives}.
\newblock
\newblock
\urldef\tempurl%
\url{https://web.archive.org/web/20180919195610/'-'
  https://sigchi.org/2018/09/sigchi-ec-values-and-strategic-initiatives/}
\showURL{%
Retrieved September 18, 2018 from \tempurl}


\bibitem[\protect\citeauthoryear{Silberman, Nathan, Knowles, Bendor, Clear,
  H{\aa}kansson, Dillahunt, and Mankoff}{Silberman et~al\mbox{.}}{2014}]%
        {Silberman2014}
\bibfield{author}{\bibinfo{person}{M.~Six Silberman}, \bibinfo{person}{Lisa
  Nathan}, \bibinfo{person}{Bran Knowles}, \bibinfo{person}{Roy Bendor},
  \bibinfo{person}{Adrian Clear}, \bibinfo{person}{Maria H{\aa}kansson},
  \bibinfo{person}{Tawanna Dillahunt}, {and} \bibinfo{person}{Jennifer
  Mankoff}.} \bibinfo{year}{2014}\natexlab{}.
\newblock \showarticletitle{Next Steps for Sustainable HCI}.
\newblock \bibinfo{journal}{\emph{interactions}} \bibinfo{volume}{21},
  \bibinfo{number}{5} (\bibinfo{date}{Sept.} \bibinfo{year}{2014}),
  \bibinfo{pages}{66--69}.
\newblock
\showISSN{1072-5520}
\urldef\tempurl%
\url{https://doi.org/10.1145/2651820}
\showDOI{\tempurl}


\bibitem[\protect\citeauthoryear{Skold and Larsen}{Skold and Larsen}{2018}]%
        {skjolddesign}
\bibfield{author}{\bibinfo{person}{Else Skold} {and} \bibinfo{person}{Frederik
  Larsen}.} \bibinfo{year}{2018}\natexlab{}.
\newblock \showarticletitle{Design for Profit or Prosperity?}
\newblock \bibinfo{journal}{\emph{Proceedings of DRS2018}}
  (\bibinfo{year}{2018}), \bibinfo{pages}{158}.
\newblock


\bibitem[\protect\citeauthoryear{Smyth and Dimond}{Smyth and Dimond}{2014}]%
        {Smyth2014}
\bibfield{author}{\bibinfo{person}{Thomas Smyth} {and} \bibinfo{person}{Jill
  Dimond}.} \bibinfo{year}{2014}\natexlab{}.
\newblock \showarticletitle{Anti-oppressive Design}.
\newblock \bibinfo{journal}{\emph{interactions}} \bibinfo{volume}{21},
  \bibinfo{number}{6} (\bibinfo{date}{Oct.} \bibinfo{year}{2014}),
  \bibinfo{pages}{68--71}.
\newblock
\showISSN{1072-5520}
\urldef\tempurl%
\url{https://doi.org/10.1145/2668969}
\showDOI{\tempurl}


\bibitem[\protect\citeauthoryear{Ssozi-Mugarura, Reitmaier, Venter, and
  Blake}{Ssozi-Mugarura et~al\mbox{.}}{2016}]%
        {SsoziMugarura2016}
\bibfield{author}{\bibinfo{person}{Fiona Ssozi-Mugarura},
  \bibinfo{person}{Thomas Reitmaier}, \bibinfo{person}{Anja Venter}, {and}
  \bibinfo{person}{Edwin Blake}.} \bibinfo{year}{2016}\natexlab{}.
\newblock \showarticletitle{Enough with 'In-The-Wild'}. In
  \bibinfo{booktitle}{\emph{Proceedings of the First African Conference on
  Human Computer Interaction}} \emph{(\bibinfo{series}{AfriCHI'16})}.
  \bibinfo{publisher}{ACM}, \bibinfo{address}{New York, NY, USA},
  \bibinfo{pages}{182--186}.
\newblock
\showISBNx{978-1-4503-4830-0}
\urldef\tempurl%
\url{https://doi.org/10.1145/2998581.2998601}
\showDOI{\tempurl}


\bibitem[\protect\citeauthoryear{Suissa}{Suissa}{2018}]%
        {Suissa2018}
\bibfield{author}{\bibinfo{person}{Judith Suissa}.}
  \bibinfo{year}{2018}\natexlab{}.
\newblock \showarticletitle{{Anarchism and Education: A philosophical
  perspective}}.
\newblock  (\bibinfo{date}{Sept.} \bibinfo{year}{2018}),
  \bibinfo{pages}{1--252}.
\newblock


\bibitem[\protect\citeauthoryear{Sun}{Sun}{2013}]%
        {Sun2013}
\bibfield{author}{\bibinfo{person}{Huatong Sun}.}
  \bibinfo{year}{2013}\natexlab{}.
\newblock \showarticletitle{{Critical Design Sensibility in Postcolonial
  Conditions}}.
\newblock \bibinfo{journal}{\emph{AoIR Selected Papers of Internet Research}}
  \bibinfo{volume}{3}, \bibinfo{number}{0} (\bibinfo{date}{Oct.}
  \bibinfo{year}{2013}).
\newblock


\bibitem[\protect\citeauthoryear{Terveen, Mentis, Quigley, and
  Palanque}{Terveen et~al\mbox{.}}{2018}]%
        {Terveen2018}
\bibfield{author}{\bibinfo{person}{Loren Terveen}, \bibinfo{person}{Helena
  Mentis}, \bibinfo{person}{Aaron Quigley}, {and} \bibinfo{person}{Philippe
  Palanque}.} \bibinfo{year}{2018}\natexlab{}.
\newblock \showarticletitle{{The evolution of SIGCHI conferences and the future
  of CHI}}.
\newblock \bibinfo{journal}{\emph{Interactions}} \bibinfo{volume}{25},
  \bibinfo{number}{5} (\bibinfo{date}{Aug.} \bibinfo{year}{2018}),
  \bibinfo{pages}{84--85}.
\newblock


\bibitem[\protect\citeauthoryear{Vaidhyanathan}{Vaidhyanathan}{2012}]%
        {Vaidhyanathan2012}
\bibfield{author}{\bibinfo{person}{Siva Vaidhyanathan}.}
  \bibinfo{year}{2012}\natexlab{}.
\newblock \bibinfo{booktitle}{\emph{{The Googlization of Everything}}}.
\newblock \bibinfo{publisher}{University of California Press}.
\newblock


\bibitem[\protect\citeauthoryear{van~der Walt}{van~der Walt}{2005}]%
        {Van2001}
\bibfield{author}{\bibinfo{person}{Lucien van~der Walt}.}
  \bibinfo{year}{2005}\natexlab{}.
\newblock \showarticletitle{Towards a history of anarchist anti-imperialism}.
\newblock \bibinfo{journal}{\emph{Against the War and Terrorism}}
  (\bibinfo{year}{2005}).
\newblock


\bibitem[\protect\citeauthoryear{Verbeek}{Verbeek}{2011}]%
        {verbeek2011moralizing}
\bibfield{author}{\bibinfo{person}{Peter-Paul Verbeek}.}
  \bibinfo{year}{2011}\natexlab{}.
\newblock \bibinfo{booktitle}{\emph{{Moralizing technology: Understanding and
  Designing the Morality of Things}}}.
\newblock \bibinfo{publisher}{University of Chicago Press}.
\newblock


\bibitem[\protect\citeauthoryear{Vorvoreanu, Zhang, Huang, Hilderbrand,
  Steine-Hanson, and Burnett}{Vorvoreanu et~al\mbox{.}}{2019}]%
        {B2019}
\bibfield{author}{\bibinfo{person}{Mihaela Vorvoreanu}, \bibinfo{person}{Lingyi
  Zhang}, \bibinfo{person}{Yun-Han Huang}, \bibinfo{person}{Claudia
  Hilderbrand}, \bibinfo{person}{Zoe Steine-Hanson}, {and}
  \bibinfo{person}{Margaret Burnett}.} \bibinfo{year}{2019}\natexlab{}.
\newblock \showarticletitle{rom Gender Biases to Gender-Inclusive Design: An
  Empirical Investigation}. In \bibinfo{booktitle}{\emph{Proceedings of the
  2019 CHI Conference on Human Factors in Computing Systems}}
  \emph{(\bibinfo{series}{CHI '19})}. \bibinfo{publisher}{ACM},
  \bibinfo{address}{New York, NY, USA}.
\newblock
\urldef\tempurl%
\url{https://doi.org/10.1145/3290605.3300283}
\showDOI{\tempurl}


\bibitem[\protect\citeauthoryear{White}{White}{2004}]%
        {White2004}
\bibfield{author}{\bibinfo{person}{Robert White}.}
  \bibinfo{year}{2004}\natexlab{}.
\newblock \bibinfo{title}{{Post Colonial Anarchism: Essays on Race, Repression
  and Culture in Communities of Color, 1999{\textendash}2004}}.
\newblock


\bibitem[\protect\citeauthoryear{White and Williams}{White and
  Williams}{2012}]%
        {White2012}
\bibfield{author}{\bibinfo{person}{Richard~J White} {and}
  \bibinfo{person}{Colin~C Williams}.} \bibinfo{year}{2012}\natexlab{}.
\newblock \showarticletitle{{The Pervasive Nature of Heterodox Economic Spaces
  at a Time of Neoliberal Crisis: Towards a
  {\textquotedblleft}Postneoliberal{\textquotedblright} Anarchist Future}}.
\newblock \bibinfo{journal}{\emph{Antipode}} \bibinfo{volume}{44},
  \bibinfo{number}{5} (\bibinfo{date}{July} \bibinfo{year}{2012}),
  \bibinfo{pages}{1625--1644}.
\newblock


\bibitem[\protect\citeauthoryear{Wilson}{Wilson}{2014}]%
        {Wilson2018}
\bibfield{author}{\bibinfo{person}{Matthew Wilson}.}
  \bibinfo{year}{2014}\natexlab{}.
\newblock \bibinfo{booktitle}{\emph{Rules Without Rulers: The Possibilities and
  Limits of Anarchism}}.
\newblock \bibinfo{publisher}{John Hunt Publishing}.
\newblock


\bibitem[\protect\citeauthoryear{Winner}{Winner}{1980}]%
        {Winner1980}
\bibfield{author}{\bibinfo{person}{Langdon Winner}.}
  \bibinfo{year}{1980}\natexlab{}.
\newblock \showarticletitle{{Do Artifacts Have Politics?}}
\newblock \bibinfo{journal}{\emph{Daedalus}} \bibinfo{volume}{109},
  \bibinfo{number}{1} (\bibinfo{year}{1980}), \bibinfo{pages}{121--136}.
\newblock


\bibitem[\protect\citeauthoryear{Winner}{Winner}{2010}]%
        {Winner2010}
\bibfield{author}{\bibinfo{person}{Langdon Winner}.}
  \bibinfo{year}{2010}\natexlab{}.
\newblock \bibinfo{booktitle}{\emph{{The Whale and the Reactor}}}.
\newblock \bibinfo{publisher}{University of Chicago Press}.
\newblock


\bibitem[\protect\citeauthoryear{Winschiers-Theophilus and
  Bidwell}{Winschiers-Theophilus and Bidwell}{2013}]%
        {WinschiersTheophilus2013}
\bibfield{author}{\bibinfo{person}{Heike Winschiers-Theophilus} {and}
  \bibinfo{person}{Nicola~J Bidwell}.} \bibinfo{year}{2013}\natexlab{}.
\newblock \showarticletitle{{Toward an Afro-Centric Indigenous HCI Paradigm}}.
\newblock \bibinfo{journal}{\emph{International Journal of Human-Computer
  Interaction}} \bibinfo{volume}{29}, \bibinfo{number}{4}
  (\bibinfo{date}{March} \bibinfo{year}{2013}), \bibinfo{pages}{243--255}.
\newblock


\bibitem[\protect\citeauthoryear{Wolff}{Wolff}{1970}]%
        {Wolff1970}
\bibfield{author}{\bibinfo{person}{Robert~Paul Wolff}.}
  \bibinfo{year}{1970}\natexlab{}.
\newblock \bibinfo{booktitle}{\emph{{In Defense of Anarchism}}}.
\newblock \bibinfo{publisher}{University of California Press}.
\newblock


\bibitem[\protect\citeauthoryear{World}{World}{2018}]%
        {IWW}
\bibfield{author}{\bibinfo{person}{International Workers of~the World}.}
  \bibinfo{year}{2018}\natexlab{}.
\newblock \bibinfo{title}{Preamble, Constitution and General Bylaws of the
  International Workers of the World}.
\newblock
\newblock
\urldef\tempurl%
\url{https://web.archive.org/web/20180606234641/https://www.iww.org/PDF/Constitutions/CurrentIWWConstitution.pdf}
\showURL{%
Retrieved September 21, 2018 from \tempurl}


\end{thebibliography}
